\documentclass[journal=tches,spthm]{iacrtrans}

\usepackage{cite}
\usepackage{amsmath,amssymb,amsfonts}
\usepackage{graphicx}
\usepackage{textcomp}
\usepackage{xcolor}
\usepackage{lipsum}
\usepackage{todonotes}
\usepackage[pangram]{blindtext}
\usepackage{xcolor}
\usepackage{tikz}
\usepackage{placeins}

\tikzstyle{n}= [circle, fill, minimum size=4pt,inner sep=0pt, outer sep=0pt]
\tikzstyle{mul} = [circle,draw,inner sep=-1pt]
\usetikzlibrary{fit,calc,positioning,decorations.pathreplacing,matrix}
\definecolor{lightgrey}{rgb}{0.82, 0.82, 0.82}

\usepackage{bm}
\usepackage{mathtools}
\usepackage{algorithm}
\usepackage{algpseudocode}
\usepackage{listings}
\usepackage{caption}
\usepackage{subcaption}
\usepackage{amsthm,aliascnt}
\usepackage[hidelinks]{hyperref}
\usepackage{pgfplots}
\usepackage{wrapfig}
\usepackage{multirow}
\usepackage{balance}

\pgfplotsset{compat=1.17}

\newcommand\Mod[1]{\enskip(\mathrm{mod}\ #1)}
\newcommand\ring{\mathcal{R}}
\newcommand\bigo{\mathcal{O}}
\newcommand\ringelement[1]{\bm{#1}}
\newcommand\ringrowvector[1]{\mathsf{#1}}

\newcommand\ringmatrixsub[2]{\overrightarrow{\ringelement{#1}_{#2}}}

\newcommand\var{X}
\DeclarePairedDelimiter\reduce{[}{]}

\DeclarePairedDelimiter\round{\lfloor}{\rceil}

\newcommand\xor{\oplus}

\algrenewcommand\algorithmicrequire{\textbf{Input:}}
\algrenewcommand\algorithmicensure{\textbf{Output:}}

\newcommand{\ceil}[1]{\left\lceil #1 \right\rceil}

\newcommand{\BASALISC}{BASALISC}

\author{Robin Geelen\inst{1}\thanks{R. Geelen and M. Van Beirendonck contributed equally to this research.} \and Michiel Van Beirendonck\inst{1}$^*$ \and Hilder V. L. Pereira\inst{1} \and Brian Huffman\inst{2} \and Tynan McAuley\inst{3} \and Ben Selfridge\inst{2} \and Daniel Wagner\inst{2} \and Georgios Dimou\inst{3} \and Ingrid Verbauwhede\inst{1} \and Frederik Vercauteren\inst{1} \and David W. Archer\inst{2}}
\institute{
  COSIC KU Leuven, Leuven, Belgium \\
  \email{firstname.lastname@esat.kuleuven.be}
  \and
  Galois, Inc., Portland, OR, USA \\
  \email{huffman@galois.com} \\
  \email{benselfridge@galois.com} \\
  \email{dmwit@galois.com} \\
  \email{dwa@galois.com}
  \and
  Niobium Microsystems, Portland, OR, USA \\
  \email{firstname@niobiummicrosystems.com}
}

\authorrunning{R. Geelen and M. Van Beirendonck et al.}

\title[\BASALISC{}: Programmable Hardware Accelerator for BGV FHE]{\BASALISC{}: Programmable Hardware Accelerator for BGV Fully Homomorphic Encryption}
\subtitle{\footnotesize Distribution Statement A: Approved for Public Release, Distribution Unlimited}

\begin{document}

\maketitle


\begin{abstract}
Fully Homomorphic Encryption (FHE) allows for secure computation on encrypted data. Unfortunately, huge memory size, computational cost and bandwidth requirements limit its practicality. We present
\BASALISC{},
an architecture family of hardware accelerators that aims to substantially accelerate FHE computations in the cloud. 
\BASALISC{} is the first to implement
the BGV scheme 
with fully-packed bootstrapping -- the noise removal capability necessary %
for
arbitrary-depth computation.
It supports a customized version of bootstrapping that can be instantiated with hardware multipliers optimized for area and power.

\BASALISC{} is a three-abstraction-layer RISC architecture, designed for a 1~GHz ASIC implementation and underway toward 150mm$^2$ die tape-out in a 12nm GF process. \BASALISC{}'s four-layer memory hierarchy includes a two-dimensional conflict-free inner memory layer that enables 32 Tb/s radix-256 NTT computations without pipeline stalls. Its conflict-resolution permutation hardware is generalized and re-used to compute BGV automorphisms without throughput penalty. 
\BASALISC{} also has a custom multiply-accumulate unit 
to accelerate BGV key switching. %

The \BASALISC{} toolchain comprises a custom compiler and a joint performance and correctness simulator. 
To evaluate \BASALISC{}, we study its physical realizability, emulate and formally verify its core functional units,
and we study its performance on a set of benchmarks. 
Simulation results show a speedup of more than 5,000$\times$ over HElib -- a popular software FHE library.
\end{abstract}

\section{Motivation}

Fully Homomorphic Encryption (FHE) \cite{rivest1978data,DBLP:conf/stoc/Gentry09,HomomorphicEncryptionSecurityStandard} offers the promise of confidentiality-preserving computation over sensitive data in a variety of theoretical and practical applications, ranging from new cryptographic primitives to machine learning as a service. Unfortunately, the utility of FHE is severely limited by its high memory size, memory bandwidth and high computational overhead. The typical result - computation that runs many orders of magnitude slower than insecure computation - prevents broad adoption. Although new schemes have markedly improved FHE performance 
\cite{brakerski2014leveled, cggi20, bipps22,ckks}, and highly optimized FHE libraries \cite{sealcrypto,DBLP:conf/crypto/HaleviS14,lattigo,WAHC:CJLOT20,polyakov2017palisade} are now available, FHE still remains orders of magnitude beyond acceptable performance limits for most potential applications.

In other computational domains where performance on general purpose processors is problematic, 
innovation has turned to purpose-built \emph{accelerators}, tuned to exploit 
domain-specific characteristics of computation. DSP accelerators, 
arguably starting with the Texas Instruments TMS320 DSP family \cite{1458134} in 1983, are 
perhaps the first example of this approach. More recently, Graphics Processing Units (GPUs) have become popular for 
accelerating video stream processing and hash function computation. Our FHE accelerator, \BASALISC{}, follows this
approach in pursuit of bringing the throughput of FHE 
computation within an order of magnitude relative to cleartext computation \cite{rondeau2020data}.

\paragraph{Contributions.}
We summarize the contributions of \BASALISC{} as follows:
\begin{itemize}
    \item \BASALISC{} accelerates BGV arithmetic for a large range of parameters. In contrast to prior accelerators, \BASALISC{} is the first to support and implement fully-packed BGV bootstrapping directly in hardware to enable unlimited-depth FHE computations.
    We propose a novel version of bootstrapping that is compatible with NTT-friendly primes. In contrast to prior work, \BASALISC{} instantiates its multipliers exclusively to these NTT-friendly primes, which saves 46\% logic area and 40\% power consumption.
    \item \BASALISC{} implements a massively parallel radix-256 NTT architecture, using a conflict-free layout, a corresponding layout permutation unit, and a twiddle factor generator. These units are deeply interleaved with the on-chip memory and provide a massive 32 Tb/s NTT throughput. In addition, we show that one can efficiently generalize the required layout permutation unit to compute BGV automorphisms without additional silicon area.
    \item \BASALISC{} is a comprehensive RISC-like architecture with a three-level Instruction Set Architecture (ISA) that allows for reasoning at diverse levels of executive abstraction. It adopts a four-level memory hierarchy purpose-built to address common FHE memory bottlenecks, including a mid-level 64~MB on-chip Ciphertext Buffer (CTB). At the lowest level, a massively parallel multiply-accumulate unit with integrated 16-entry register file allows accelerating tight BGV key switching loops, asynchronously and independently of the CTB.
    \item \BASALISC{} is placed and routed with 150mm$^2$ die size and 1~GHz operational frequency in a 12nm low-power Global Foundries process. Critical hardware logic is emulated and formally verified for correctness. \BASALISC{} is evaluated on a set of micro- and macro-benchmarks, showing more than %
    5,000$\times$ speedup over HElib.

\end{itemize}

\section{Preliminaries}

\subsection{Fully Homomorphic Encryption}

Fully Homomorphic Encryption (FHE) provides a simple use model to securely outsource computation on sensitive data to a 
third party. Informally, the FHE model enables a user to encrypt their data $m$ into a ciphertext $c = \text{Enc}(m)$, then send it
to a third party, who can compute on $c$. The third party produces another ciphertext $c'$
encrypting $f(m)$ for some desired function $f$.
This is done by representing the function in terms of the operations provided by the scheme, typically addition and multiplication,
and computing these operations on the encrypted data.
We say that $f$ was computed homomorphically.

\begin{wrapfigure}{r}{0.37\linewidth}
    \centering
    \includegraphics[width=\linewidth]{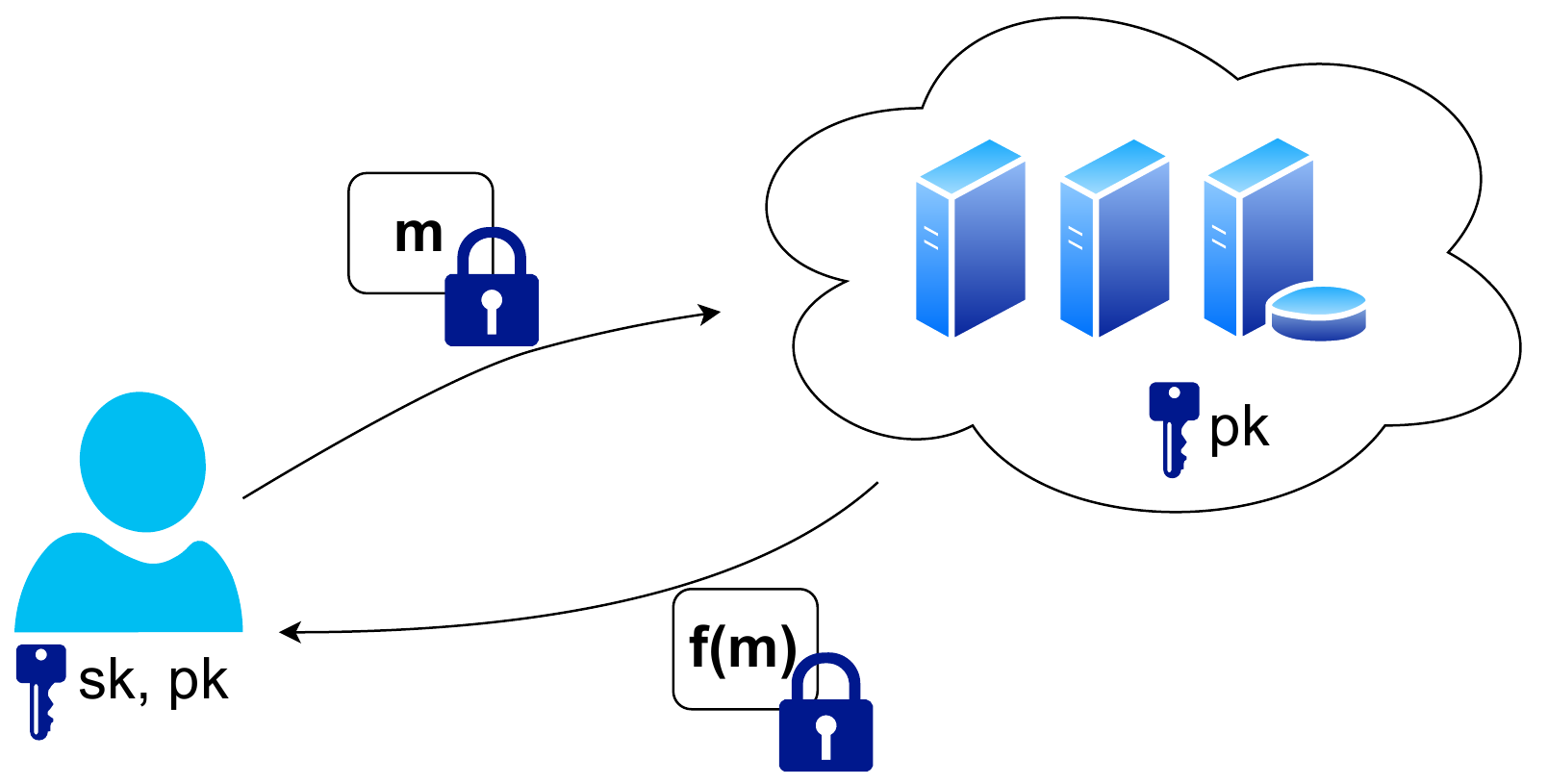}
    \caption{FHE used in a typical commercial application.}
    \label{fig:FHE}
\end{wrapfigure}

In FHE, the third party receives only ciphertexts and the public key, but
never the secret key that allows decryption. As a result, the sensitive inputs are protected under the security of the encryption scheme.
Because the result of the computation remains encrypted, the output also remains 
unknown to the third party: only the holder of the secret key can decrypt and access it. This scenario is illustrated in \autoref{fig:FHE}.

To achieve security, the ciphertexts of all FHE schemes are noisy. This noise is added to the input data during encryption and removed during decryption.
Each homomorphic operation, such as an addition or a multiplication, increases the noise in the resulting ciphertext.
Decryption can still recover the correct result, provided that the noise is small enough.
Therefore, we can compute only a limited number of homomorphic operations before we reach the
limit of decryption failure.

Because multiplications increase ciphertext noise much more than additions, we usually model noise growth
by the number of
sequential multiplications.
Computing the product $\prod_{i=1}^L m_i$ %
requires a \emph{multiplicative depth} of %
$\ceil{\log_2(L)}$. This is accomplished by writing the product in a tree structure, with each leaf node representing one of the factors.
Note that FHE suffers from a general trade-off between 
computational cost
and 
tolerating a larger $L$: one can increase the encryption parameters
as to %
obtain more multiplicative depth,
but in doing so, the homomorphic operations become slower
and the size of ciphertexts larger.

To support the computation of functions regardless of their multiplicative depth,
FHE uses \emph{bootstrapping}. This operation reduces noise by decrypting a ciphertext homomorphically.
Unfortunately, bootstrapping is very expensive, so its use is often minimized. %
There are several techniques in the FHE literature to slow down the noise growth, and thus delay bootstrapping.
This work employs \emph{key switching} and 
\emph{modulus switching}~\cite{brakerski2014leveled}.
In practice, bootstrapping and key switching
tend to heavily 
dominate computation and data movement costs of an 
application: 
in a simple 1,024-point, 10-feature logistic regression, we see that
these tasks 
account for over 95\% of the computational effort and the vast majority of data 
movement.

In summary, 
the implementation challenges of FHE are the high complexity of computation, the large ciphertext expansion ratio (large polynomials with integer coefficients of a 1000 bits or more), and the proportion of effort needed in 
bootstrapping (or delaying it) in sufficiently complex programs. 
In the remainder of this paper, we examine the magnitude of these challenges and how they 
impact the design of our FHE accelerator.

\subsection{The BGV Cryptosystem}
\BASALISC{} targets the homomorphic encryption scheme %
BGV~\cite{brakerski2014leveled}.
Plaintexts and ciphertexts are represented by elements in the ring $\ring = \mathbb{Z}[\var]/(\var^N+1)$ with $N$ a power of~2. Those elements are thus polynomials reduced modulo $\var^N+1$, which %
is implicit in our notation. BGV guarantees finite data structures by additionally reducing the coefficients: the plaintext space is computed modulo $t$ (denoted $\ring_t$), and the ciphertext space is a pair of elements modulo $q$ (denoted $\ring_q^2$).
Reduction modulo $m$ (with $m = t \text{ or } q$) is explicitly denoted by $\reduce{\cdot}_{m}$ and is done symmetrically around 0 (i.e., in the set $[-m/2, m/2) \cap \mathbb{Z}$).

As with traditional ciphers, BGV has encryption and decryption procedures to move between the \emph{plaintext space} and the \emph{ciphertext space}. These operations are never executed by the server that performs the outsourced computation, and are therefore not implemented by \BASALISC{}. However, the ciphertext format remains an important element of the BGV scheme and is essential to explain the homomorphic operations. A BGV ciphertext $(\ringelement{c}_0, \ringelement{c}_1) \in \ring_q^2$ is said to encrypt plaintext $\ringelement{m} \in \ring_t$ under secret key $\ringelement{s}$ (which has small coefficients) if
\begin{equation}
\label{eq:decryption}
\ringelement{c}_0 + \ringelement{c}_1 \cdot \ringelement{s} = \ringelement{m} + t\ringelement{e} \Mod{q}
\end{equation}
for some element $\ringelement{e}$ that also has small coefficients. The term $\ringelement{e}$ is called the \emph{noise}, and it determines if decryption returns the correct plaintext: if $\ringelement{e}$ has coefficients roughly less than $q/2t$, then $\ringelement{m} + t\ringelement{e}$ does not overflow modulo $q$. We can therefore recover the plaintext uniquely as $\ringelement{m} = \reduce{\reduce{\ringelement{c}_0 + \ringelement{c}_1 \cdot \ringelement{s}}_q}_t$.

\subsubsection{Basic Homomorphic Operations}
\label{sec:basicops}

Smart and Vercauteren~\cite{smart2014fully} observed that for $t = p^r$ with $p$ an odd prime, the plaintext space $\ring_t$ is equivalent to $\mathbb{Z}_t^\ell$ for some $\ell$ that divides $N$. This technique - referred to as \emph{packing} - allows us to encode $\ell$ numbers into one plaintext simultaneously. Addition and multiplication over tuples in $\mathbb{Z}_t^\ell$ are then performed entry-wise. As a result, one ciphertext can encrypt and operate on an entire tuple, which leads to significant performance benefits and memory reductions in practice.

When BGV is used in conjunction with packing, one can define
three basic homomorphic operations.
Let
$(\ringelement{c}_0, \ringelement{c}_1)$ and $(\ringelement{c}'_0, \ringelement{c}'_1)$
be two distinct ciphertexts that encrypt respectively the tuples
$(m_1, ..., m_\ell)$
and 
$(m_1', ..., m_\ell')$, then we can perform the following operations:

\begin{itemize}
    \item \textbf{Addition:}
    compute $\ringrowvector{ct}_{\text{add}} = (\reduce{\ringelement{c}_0 + \ringelement{c}'_0}_q, \reduce{\ringelement{c}_1 + \ringelement{c}'_1}_q)$. 
    The underlying plaintext of $\ringrowvector{ct}_{\text{add}}$
    is then
    $(m_1 + m_1', ..., m_\ell + m_\ell')$.
    \item \textbf{Multiplication:} compute 
    $\ringrowvector{ct}_{\text{mul}} = (\reduce{\ringelement{c}_0 \cdot \ringelement{c}'_0}_q,
    \reduce{\ringelement{c}_0 \cdot \ringelement{c}'_1 +
    \ringelement{c}_1 \cdot \ringelement{c}'_0}_q, \reduce{\ringelement{c}_1 \cdot \ringelement{c}'_1}_q)$. 
    Observe that the resulting ciphertext consists of three ring elements, but this can be reduced back to two with a post-processing step known as \emph{key switching}.
    The underlying plaintext of $\ringrowvector{ct}_{\text{mul}}$ is then
    $(m_1 \cdot m_1', ..., m_\ell \cdot m_\ell')$.

    \item \textbf{Permutation:} compute 
    $\ringrowvector{ct}_{\text{per}} = (\phi_k(\ringelement{c}_0),
    \phi_k(\ringelement{c}_1))$. The
    map $\phi_k$ (called \emph{automorphism}) is parameterized by an odd integer $k$ and defined as $\phi_k \colon c(\var) \mapsto c(\var^k)$. Gentry et 
    al.~\cite{gentry2012fully} have shown that automorphisms induce a 
    permutation on the elements of the encoded tuple. The underlying plaintext of $\ringrowvector{ct}_{\text{per}}$ is therefore %
    a permutation of
    $(m_1, ..., m_\ell)$.
    Although the resulting ciphertext has only two elements, 
    we still need post-processing by means of key switching.
    
\end{itemize}
The validity of these three operations can be verified by observing their effect on \autoref{eq:decryption}.
We refer to Zucca~\cite{zucca2018towards} for more details, including an analysis of the noise growth of each homomorphic operation.

\subsubsection{Auxiliary Homomorphic Operations}
Basic homomorphic operations lead to ciphertext expansion and noise growth.
A product ciphertext, for example, consists of three elements
and is encrypted under
$(\ringelement{s}, \ringelement{s}^2)$
instead of~$\ringelement{s}$. 
Moreover, the noise term is equal to $t\ringelement{e} \cdot \ringelement{e}'$, where $\ringelement{e}$ and $\ringelement{e}'$ are the noise terms of the input ciphertexts.
A similar problem occurs during permutation, where the automorphism~$\phi_k$ not only acts on the ciphertext and plaintext, but also on %
the secret key. 
The resulting ciphertext will therefore be encrypted under 
$\phi_k(\ringelement{s})$ instead of $\ringelement{s}$.

To prevent ciphertext expansion, maintain the same encryption key for each ciphertext, %
and slow down noise growth, BGV defines two auxiliary procedures:
\begin{itemize}
    \item \textbf{Modulus switching:} given a ciphertext 
    $(\ringelement{c}_0, \ringelement{c}_1) \in \ring_q^2$ and a modulus 
    $q'$, compute a new ciphertext 
    $(\ringelement{c}_0', \ringelement{c}_1') \in \ring_{q'}^2$ that decrypts with respect to $q'$ instead of $q$.
    Modulus switching has the positive side effect of reducing the noise by a factor of $q' / q$.
    
    \item \textbf{Key switching:} given a key switching matrix $(\ringmatrixsub{k}{0}, \ringmatrixsub{k}{1})$ and either a product ciphertext $(\ringelement{c}_0, \ringelement{c}_1, \ringelement{c}_2) \in \ring_q^3$ or a permuted ciphertext $(\ringelement{c}_0, \ringelement{c}_1) \in \ring_q^2$, compute a new ciphertext $(\ringelement{c}_0', \ringelement{c}_1') \in \ring_{q}^2$ that decrypts directly under the secret key $\ringelement{s}$ using \autoref{eq:decryption}. Thus key switching brings the ciphertext back to its original format.
\end{itemize}
Modulus switching is typically done just before multiplication in order to reduce the noise to its minimum level.
Key switching is performed after multiplication or permutation 
to keep the ciphertext format consistent. We again refer to Zucca~\cite{zucca2018towards} for more details, including an analysis of the noise growth of both operations.

\subsubsection{Bootstrapping}
When the entire noise budget of a ciphertext is consumed (equivalently, when the modulus is depleted to its minimum value
by successive modulus switchings),
further homomorphic operations are no longer immediately possible. We can overcome this problem by means of a \emph{bootstrapping} procedure that reduces the noise back to a lower level~\cite{DBLP:conf/stoc/Gentry09}. Bootstrapping refreshes a ciphertext by running decryption \emph{homomorphically}. One evaluates an adapted version of \autoref{eq:decryption}, followed by a %
homomorphic
rounding procedure. The state-of-the-art bootstrapping technique for BGV is implemented in the HElib software library~\cite{halevi2021bootstrapping}.

\section{Data Representation and Algorithms}
\label{sec:data-alg}
Basic homomorphic operations can be implemented via arithmetic in $\ring_q$, i.e., based on polynomial addition, multiplication and automorphism.
In order to obtain efficient arithmetic in $\ring_q$, two common tricks have been developed~\cite{gentry2012homomorphic}, and \BASALISC{} employs them as well.
Firstly, \autoref{sec:rns} explains how computations modulo $q$ can be split into many smaller moduli~$q_i$, based on the Chinese Remainder Theorem (CRT). This data representation is called a Residue Number System (RNS). Secondly, \autoref{appendix:ntt} explains conversion between the polynomial and frequency domain via the Number-Theoretic Transform (NTT), which is necessary for efficient multiplication and automorphism.
Similarly to RNS, the NTT can also be interpreted in terms of the Chinese Remainder Theorem.
Hence, the combination of using RNS for fast arithmetic modulo $q$ and the NTT for fast polynomial arithmetic modulo $X^N+1$, is referred to as \emph{Double-CRT} representation.

\subsection{Residue Number System}
\label{sec:rns}
Suppose that the ciphertext modulus is given by the product $q = q_1 \cdot \ldots \cdot q_k$ of pairwise coprime numbers. Then %
computations in $\ring_q$ can be reduced to simultaneous computations in the smaller rings $\ring_{q_i}$ by applying the Chinese Remainder Theorem. %
This data representation is called a Residue Number System (RNS), and brings an asymptotic speedup factor of $\bigo(k)$. Moreover, it simplifies architecture design, because the size of each $q_i$ is much smaller than $q$ (a typical value is 32 bits for $q_i$ versus more than 1000 bits for $q$).

\subsection{Number-Theoretic Transform}
\label{appendix:ntt}

In order to perform efficient polynomial multiplication in time $\mathcal{O}(N\log(N))$,
we resort to the Number-Theoretic Transform~(NTT). The NTT is a generalization of the 
Fast Fourier Transform~(FFT) to finite fields, %
and allows us to use exact integer 
arithmetic, preventing round-off errors typical of real-valued FFT computations. Similar to the FFT, the NTT can be computed with the Cooley-Tukey algorithm that recursively re-expresses an NTT of size $N = N_1N_2$ as $N_2$ inner NTTs of size $N_1$, followed by $N_1$ outer NTTs of size $N_2$. Before the outer NTT, each output of the inner NTT is multiplied by a twiddle factor:

\begin{gather}
    X[k_1{+}N_1k_2] = \sum_{n_2 = 0}^{N_2-1}\left(\sum_{n_1=0}^{N_1-1}x[N_2n_1{+}n_2]\omega_{N_1}^{n_1k_1} \right)\omega_N^{n_2k_1} \omega_{N_2}^{n_2k_2}.
   \label{eq:cooley-tukey}
\end{gather}

\noindent By choosing $N_1 = 2$ and $N_2 = N/2$ at each recursive decomposition or vice-versa, the well-known radix-two Decimation-In-Time (DIT) and Decimation-In-Frequency (DIF) algorithms are obtained, respectively.

The NTT can be used directly for fast cyclic convolutions (polynomial multiplication modulo $X^N-1$). However, BGV %
performs polynomial multiplication modulo $X^N+1$,
requiring \emph{negacyclic} convolutions. Those %
can still be implemented with a regular NTT, but require an additional pre-multiplication of the
input polynomials and post-multiplication of the output polynomial by an extra set of twiddle factors~\cite{DBLP:books/aw/AhoHU74}.

\subsection{Implemented Algorithms and Parameter Sets}
\BASALISC{} accelerates the five basic and auxiliary homomorphic operations. All of these can be built from three building blocks on vector operands: entry-wise addition and multiplication; entry permutation; and the NTT.
For example, addition is coefficient-wise and handled directly in either the polynomial or frequency domain. %
On the other hand, multiplication and automorphism can only be handled in the frequency domain, respectively via entry-wise multiplication and entry permutation.
Modulus and key switching are more complicated, and they need conversion between the polynomial and frequency domain via the NTT.

\subsubsection{Supported Parameter Sets}
As opposed to software implementations, %
hardware accelerators %
gain throughput benefits by supporting a limited range of commonly used parameters.
We start with the realization that at least 128-bit security must be supported if \BASALISC{} is to be interesting to real-world users.
Based on this observation, we choose a parameter range that allows for an efficient implementation, while still retaining sufficient freedom for application design.
A typical range for the ring dimension $N$, %
offering sufficient flexibility, is between $2^{14}$ and $2^{16}$. \BASALISC{} settles on a maximum value of $N = 2^{16}$, which allows ciphertext moduli up to $q = 2^{1782}$ at 128-bit security level. This gives a large number of multiplicative levels, even at a high-precision plaintext space (e.g., 31 levels at plaintext modulus $t = 127^3$ without bootstrapping; with bootstrapping, we get an arbitrary number of levels).

\autoref{tab:params} shows the full parameter range supported by \BASALISC{} and an example parameter set for illustration. The largest ciphertext modulus  that appears during the basic homomorphic operations is denoted by $Q$; during key switching, however, the modulus is temporarily extended to $QP$. %
Concretely, our largest supported modulus is $QP \approx 2^{1782}$.

\begin{table}[b]
    \centering
    \caption{\BASALISC{} parameter ranges and examples.}
    \label{tab:params}
    \small
    \vspace*{3pt}
    \begin{tabular}{c|c|c}
        Parameter & Range & Example \\ \hline
        Security parameter & N/A & 128 bits \\
        Ring dimension $N$ & $512-65536$ & 65536 \\
        Plaintext modulus $t$ & $> 2$ & $127^3$ \\ %
        Ciphertext packing $\ell$ & $2-65536$ & 64 slots \\
        Max $\log_2(QP)$ for key switching & $20 - 1782$ & 1782 bits \\
        Max $\log_2(Q)$ for ciphertext & $20 - 1782$ & 1263 bits \\
        Max multiplicative depth $L$ & N/A & 31 
    \end{tabular}
    
\end{table}

\subsubsection{Algorithmic Details}

Currently, the term BGV refers to a family of related algorithms that are derived from the original scheme~\cite{brakerski2014leveled}. One assumption of the original scheme is that the factors of the ciphertext modulus satisfy $q_i = 1 \Mod{t}$. This can be interpreted as a $\log_2(t)$-bit restriction on $q_i$, which limits the freedom in the selection of the ciphertext modulus and poses a lower bound on the factors of the RNS chain. HElib implements an improved variant of the BGV scheme that does not require this restriction~\cite{halevi2020design}. However, this improved %
version
needs to keep track of a \emph{correction factor} $\kappa \in \mathbb{Z}^*_t$ and encrypts $\kappa \cdot \ringelement{m}$ instead of $\ringelement{m}$. Each ciphertext is tagged with such a correction factor, which is %
removed upon decryption by multiplying with $\kappa^{-1}$.
\BASALISC{} implements the HElib version of BGV. Correction factor management is done by our compiler Artemidorus (see \autoref{sec:compileandsim}).

Due to
algebraic constraints, the NTT is only defined for prime moduli %
of the shape
$q_i = 1 \Mod{2N}$. These special moduli are referred to as \emph{NTT-friendly primes}.
In the case of our ring dimension $N = 2^{16}$, this puts a lower bound of 17 bits on the size of~$q_i$. Coupled with the requirement to have a sufficient amount of
moduli available to reach $\log_2(QP) = 1782$ bits, a simple analysis can show that we need $q_i$ of at least 26 bits. %
In practice, however, %
\BASALISC{} employs 32-bit moduli, because it gives a better utilization for the on-chip memory buffer and simplified interaction with the external memory. Furthermore, %
both 26-bit and 32-bit moduli result in the same complement of arithmetic units within our silicon area budget. For the example parameter set of \autoref{tab:params}, 
$Q$ is a product of 42 primes and 
$P$ is a product of 14 additional primes.

\section{NTT-Friendly Bootstrapping}
\label{sec:harwareBootstrap}
Since the number-theoretic transform is only defined for NTT-friendly primes, we restrict \BASALISC{}'s hardware multipliers to these NTT-friendly primes (see \autoref{sec:mont-mul-design} for the motivation of this design choice).
However, all current bootstrapping implementations switch to a specially-shaped modulus~\cite{halevi2021bootstrapping,chen2018homomorphic}, either $q = p^e + 1$ or $q = p^e$, which is not an NTT-friendly prime in general.
For software implementations, this is not an issue, because CPUs provide a general-purpose instruction set.
However, with our specialized hardware multipliers, bootstrapping cannot be implemented directly.

We propose a generalized version of bootstrapping that works with NTT-friendly primes exclusively, and can still be evaluated with the same computational cost as regular bootstrapping. %
It relies on a simplified decryption lemma that was proposed earlier~\cite{geelen2022bootstrapping}.
Our contribution is to implement this lemma in \BASALISC{}, which requires non-trivial RNS operations as explained in the remainder of this section.

\begin{lemma}[Simplified decryption~\cite{geelen2022bootstrapping}]
\label{lem:inner-product}
Let $p$ be a prime number, and let $e > r \geqslant 1$ and~$q$ be sufficiently high parameters with $\gcd(q, p)=1$. If $(\ringelement{c}_0, \ringelement{c}_1)$ is a BGV encryption of $\ringelement{m}$ under plaintext modulus $p^r$ and ciphertext modulus $q$, then it can be decrypted as
\[\ringelement{c}_i' \gets \reduce{p^{e - r} \ringelement{c}_i}_{q},\quad\ringelement{w} \gets \reduce{q^{-1}\cdot(\ringelement{c}_0' + \ringelement{c}_1' \cdot \ringelement{s})}_{p^e}\quad\text{and}\quad\ringelement{m} \gets \reduce{q\cdot\round{\ringelement{w}/p^{e-r}}}_{p^{r}},\]
where $q^{-1} \cdot q = 1 \Mod{p^e}$, and $\round{\cdot}$ denotes coefficient-wise rounding to the nearest integer.
\end{lemma}

Our NTT-friendly bootstrapping evaluates \autoref{lem:inner-product} in the homomorphic domain, and it uses
two primitives from Bajard et al.~\cite{bajard2016full}. %
Consider two coprime moduli
\begin{equation*}
Q = \prod_{i=1}^k q_i\qquad\text{and}\qquad P = \prod_{j=1}^\ell p_{j}.
\end{equation*}
The first primitive - \emph{fast base extension} - extends the modulus from $Q$ to $QP$ and is defined as follows. 
Given an element $\ringelement a \in \ring_{Q}$ 
in polynomial representation, compute 
\begin{equation}\label{eq:fast-base-extension}
\Call{FastBaseExt}{}_{Q\to QP} (\ringelement a) = \left(\reduce*{\sum_{i=1}^k
                    \reduce*{ \ringelement a \cdot \left(\frac{Q}{q_i}\right)^{-1} }_{q_i}
                    \cdot \frac{Q}{q_i}}_{p_j}  \right)_{1\leqslant j\leqslant \ell}\in\ring_{QP}.
\end{equation}
Fast base extension possibly incurs additional overflows modulo $Q$. This means that the output will be equal to $\ringelement{a} + Q\ringelement{r}$ for some $\ringelement{r}\in \ring$. However, denoting the infinity norm by $||\cdot||_{\infty}$, the overflows will be upper bounded as $||\ringelement{r}||_{\infty}\leqslant k/2$.

The second primitive - \emph{small Montgomery reduction} - 
takes an element $\ringelement{b} \in \ring_{QP}$ in polynomial representation subject to $||\ringelement{b}||_{\infty}\ll P\cdot m$, and it outputs $\ringelement{c} \in \ring_Q$ such that $\ringelement{c} = \ringelement{b}\cdot P^{-1}\Mod{m}$. Moreover, the coefficients of the result will be reduced modulo $m$, i.e., they are upper bounded as $||\ringelement{c}||_{\infty} \leqslant (1 + \epsilon)m/2$ for some $\epsilon \ll 1$. %
This functionality is denoted by $\Call{SmallMont}{}_{QP\to Q}(\ringelement{b}, m)$.
During bootstrapping, we use small Montgomery reduction to compensate for the overflows of fast base extension and to compute reduction modulo $p^e$ as required in the decryption formula.

NTT-friendly bootstrapping is specified in \autoref{alg:bootstrapping} and is a translation of \autoref{lem:inner-product} to the homomorphic domain.
We use three pairwise coprime moduli $Q$, $q$ and $b$, where the first two are products of NTT-friendly primes, and the last one is an NTT-friendly prime.
Conversion between a ciphertext and its ring elements is implicit in our notation: at every place in the algorithm, we have $\ringrowvector{ct} = (\ringelement{c}_0, \ringelement{c}_1)$ %
with correction factor $\kappa$
(and similarly for other ciphertexts).
Note that the input ciphertext and all other variables are assumed to be in polynomial representation.

The algorithm first multiplies the ciphertext by $p^{e-r}$ and converts to Montgomery representation with respect to the auxiliary base $b$.
Then the modulus is lifted to $Q\cdot q\cdot b$, which causes extra overflows modulo $q$. These are compensated almost perfectly by the small Montgomery reduction on the next line, which also converts back from Montgomery to regular representation (i.e., it removes the additional $b$-factor).
Then we reduce modulo~$p^e$ via another small Montgomery reduction, and as a side effect, the ciphertext gets multiplied by $q^{-1}$, which is required in the decryption formula.
The resulting ciphertext $\ringrowvector{ct}'' = (\ringelement{c}_0'', \ringelement{c}_1'')$ is an encryption of $\ringelement{w}$ under plaintext modulus $p^e$ and ciphertext modulus~$Q$. Observe that taking the product with the secret key $\ringelement{s}$ is implicit and does not require any computation during bootstrapping~\cite{alperin2013practical}.
Finally, we perform the homomorphic rounding procedure in the same way as HElib and restore the original correction factor. The correction factor is also augmented with an additional factor of $q^{-1}$, which represents the multiplication by~$q$ that is required in the decryption formula.
Note that our algorithm does not change key generation nor encryption, so it does not influence the security of BGV.

\begin{algorithm}[H]
\small
\caption{NTT-friendly bootstrapping}
\label{alg:bootstrapping}
\begin{algorithmic}[1]
\Require $\ringrowvector{ct} \in \ring_q^2$ with noise $\ringelement{e}$ %
\Ensure $\ringrowvector{ct}'' \in \ring_{q''}^2$ with noise $\ringelement{e}''$ 
s.t. $\text{Dec}(\ringrowvector{ct}'') = \text{Dec}(\ringrowvector{ct})$ and $||\ringelement{e}''||_{\infty}/q'' \ll ||\ringelement{e}||_{\infty}/q$
\Function{Bootstrap}{}($\ringrowvector{ct}$) \Comment{Store correction factor $\kappa$ of $\ringrowvector{ct}$}

\For{$i \in \{0, 1\}$}

\State $\ringelement{c}_i' \gets \reduce{\ringelement{c}_i \cdot p^{e-r} \cdot b}_q$ \label{line:step1} \Comment{Residues defined mod $q$}
\State $\ringelement{c}_i' \gets$ \Call{FastBaseExt$_{q\to Q\cdot q\cdot b}$}{$\ringelement{c}_i'$} \label{line:step2} \Comment{Residues defined mod $Q \cdot q \cdot b$}
\State $\ringelement{c}_i' \gets$ \Call{SmallMont$_{Q\cdot q \cdot b\to Q\cdot q}$}{$\ringelement{c}_i', q$} \label{line:step3} \Comment{Reduce mod $q$ and drop $b$}
\State $\ringelement{c}_i'' \gets$ \Call{SmallMont$_{Q\cdot q \to Q}$}{$\ringelement{c}_i', p^e$} \label{line:step4} \Comment{Reduce mod $p^e$ and drop $q$}

\EndFor

\State $\ringrowvector{ct}'' \gets \round{\ringrowvector{ct}'' / p^{e-r}}$ \label{line:step6} \Comment{Same as in HElib with initial $\kappa'' \gets 1$}
\State \Return $\ringrowvector{ct}''$ \Comment{Update correction factor as $\kappa'' \gets \kappa'' \cdot \kappa \cdot q^{-1}$}
\EndFunction
\end{algorithmic}
\end{algorithm}

\section{\BASALISC{} Instruction Set Architecture}

\BASALISC{} is an adapted Reduced Instruction Set Computer (RISC) architecture with a three-level instruction set. This multi-level approach allows for reasoning at diverse levels of abstraction, and aids in assuring correctness of our system. Having a hierarchy of multiple intermediate representations and instruction sets, each with well-defined semantics, means that we can implement and test each stage of the compiler toolchain separately. In addition, different instruction set abstractions allow programmers to work at a higher level of abstraction while allowing compiler writers and library authors to reason about lower-level details such as scheduling and optimizations easily. For example, when writing a program to run on \BASALISC{}, the programmer need not know about low-level data representations. 
We generate and reason over three distinct levels of instruction and typesystem abstraction:
\begin{itemize}
    \item \textbf{Macro-instructions} are at the highest level, with the largest data types and the most complex operations. Entire ciphertexts, plaintexts, and key switching matrices are treated as basic data types
    at this abstraction level.
    Operations
    include ciphertext addition, multiplication, modulus and key switching, 
    automorphisms, and bootstrapping. Details about data representation and algorithms that implement those operations are opaque at this level of abstraction.
    \item \textbf{Mid-level instructions} expose the Double-CRT data representation. The basic data type at this level is a residue polynomial (a polynomial in RNS representation) comprising up to $2^{16}$ 32-bit polynomial coefficients. %
    Basic operations on these data types include pointwise modular addition and multiplication on vectors of coefficients; automorphisms;  NTTs; and multiply-accumulate iterations commonly used in key switching. Also included in this list are memory management instructions that Load and Store data to and from off-chip memory.
    \item \textbf{Micro-instructions} correspond very closely with the specific operations performed by the processing elements (PEs). The basic data type at this level contains as many coefficient words (1024 or 2048) as can be processed simultaneously by a PE or accessed in one on-chip memory cycle. Instructions at this level are delivered via the Peripheral Component Interconnect Express (PCIe) interface to the \BASALISC{} processor for execution. 
    This instruction level also includes rudimentary machine control instructions.
\end{itemize}

The left subtable of \autoref{tab:opcodes-modes} shows operation code mnemonics (opcodes) from each level of our instruction set. Together with opcodes, \BASALISC{} uses operand specifiers (omitted from the table) with register-like addressing modes for all levels in its ISA and memory hierarchy.
The right subtable of \autoref{tab:opcodes-modes}
shows addressing mode examples for operand specifiers at the Micro-instruction level. At this level, operands are either ``chunks'' of 2048 32-bit coefficients within a residue polynomial, 32-bit scalar values, or natural number indices into tables of moduli.

\begin{table}
\centering
\caption{Example opcodes (left) and micro-level operand addressing modes (right).}
\label{tab:opcodes-modes}
\scriptsize
\vspace*{3pt}
\begin{tabular}{c|c|c}
    ISA & Opcode & Semantics \\ \hline
    Macro   & LOAD        & Move data from distant to near memory \\
            & KSW   & Key switch a ciphertext \\
            & MORPH       & Perform automorphism on a ciphertext \\ \hline
    Mid     & MULI          & Multiply a residue polynomial by constant \\
            & NTT           & Compute NTT of residue polynomial \\
            & FBE & Fast Base Extension \\ \hline 
    Micro   & NTT1          & Perform iteration of first NTT pass\\
            & MAC        & Multiply operands and add to accumulator
\end{tabular}
\quad
\centering
\scriptsize
\begin{tabular}{c|c}
    Mode  & Definition \\ \hline
    \$XXX                   & address in distant memory \\
                            & only for LOAD/STORE \\
    rXXX                    & address in middle memory \\
    tXX                     & register in near memory \\
    nXXX                    & immediate 32-bit scalar \\
    iXXX                    & index in moduli table\\
\end{tabular}
\end{table}

\section{\BASALISC{} Hardware Design}

\begin{wrapfigure}{r}{0.66\linewidth}
  \centering
  \includegraphics[width=\linewidth]{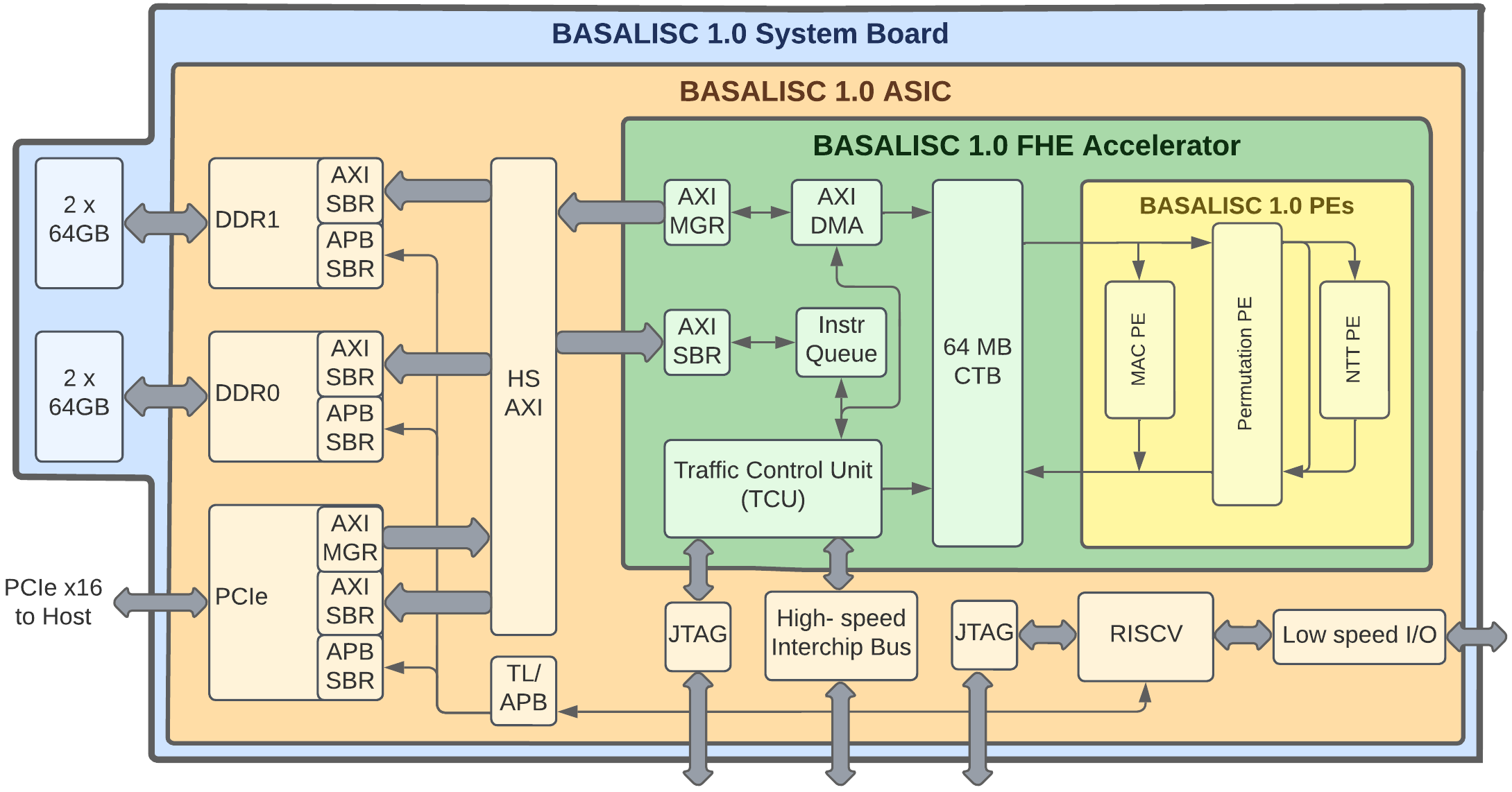}
  \caption{\BASALISC{} System Diagram.}
  \label{fig:system}
\end{wrapfigure}

\autoref{fig:system} shows a system diagram of \BASALISC{}. \BASALISC{} is a single-chip FHE coprocessor, designed in a 12nm Global Foundries process, with additional off-chip memory; high-speed connectivity to its host system; and extensibility via a high-speed inter-chip interconnect. For its implementation, \BASALISC{} employs a mature asynchronous logic process controlled by standard handshaking protocols \cite{8259423}, allowing units to accelerate to a higher clock frequency whenever
input data is available. \autoref{sec:mac-pe} describes how this design choice %
helps to accelerate costly key switchings.

\subsection{Memory Architecture}
\label{sec:memorySubsystem}

\BASALISC{} includes four layers of memory hierarchy that exhibit diverse latencies and capacities. \autoref{tab:memhier} depicts these four layers, where - typical of computer memory hierarchies - lower latency layers have smaller capacities. From farthest to nearest to the PEs, these four layers comprise off-chip distant memory (DRAM), a middle memory Ciphertext Buffer (CTB), %
a local Register File (RF) and Accumulator register (ACC) within the MAC PE. 
A significant difference between typical memory hierarchies and that of \BASALISC{} is the working set size that each layer can hold. Still, we expect capacity limits of layers in our memory hierarchy to be a major limiter of system performance. In particular, we expect minimal locality of reference for key switching matrices, each of which is larger than the entire CTB.
We now describe the DRAM array and the CTB, and defer the description of the MAC memory architecture to \autoref{sec:mac-pe}.

\begin{table}[b]
    \centering
    \caption{Memory hierarchy for ciphertext and key storage.}
    \label{tab:memhier}
    \small
    \vspace*{3pt}
    \begin{tabular}{c|r|r}
    Memory & Capacity & Round-trip latency \\ \hline
      Off-chip DRAM  & 256~GB & $>$ 100~ns  \\
      CTB     & 64~MB & $\sim$3~ns \\
      MAC RF & 128~kB & $\sim$1.25~ns \\
      MAC ACC & 8 kB & 0.625~ns 
    \end{tabular}
\end{table}

\begin{itemize}

\item \textbf{The 64~MB CTB} contains $2^{24}$ locations, each of which holds a 32-bit residue polynomial coefficient. In our largest supported parameter set, a single residue polynomial consists of $N=2^{16}=64$K coefficients and occupies one entire page of the CTB. 
The CTB is a single-port SRAM array that can either read or write 2048 32-bit residue polynomial coefficients every machine cycle, providing a total bandwidth of 8 Tb/s (at 1 GHz operation) to our set of data Processing Elements (PEs) shown in yellow. 
As an advantage of FHE determinism, allocation and size of all data and operands are bound at compile-time. This allows the \BASALISC{} CTB to be structured as an addressable set of ciphertext registers, instead of requiring the complex functionality of a run-time cache memory. This set of registers is compiler-managed with a true Least-Recently Used (LRU) replacement policy. CTB bandwidth is not materially affected by concurrent transfer between distant memory and the CTB: roughly at most 0.3\% of CTB access cycles are used by our total distant memory bandwidth. 

\item \textbf{The 256~GB DRAM} serves as the staging area for data that is scheduled for processing and for results that are ready for retrieval by the host computer. In addition, for many applications, the 64~MB CTB is too small to hold the sizeable working sets of ciphertexts and key switching matrices. The DRAM array ensures that CTB capacity misses do not have to spill to host memory.

\end{itemize}

\subsection{System Design}

In contrast to other FHE hardware accelerators \cite{10.1145/3466752.3480070}, \BASALISC{} reduces cost and manufacturing risk by relying only on commercially available standard packaging, DRAM, and PCIe technologies and fits within 150mm$^2$ \cite{rondeau2020data}. From the top down, the \BASALISC{} system can be described in different levels of hierarchy (see \autoref{fig:system}):
\begin{itemize}
    \item \textbf{(Blue) The \BASALISC{} System Board} instantiates the distant DRAM memory using two DDR4 subsystems, each providing up to 128~GB of DRAM and 25.6~GB/s of bandwidth. At bottom left of the diagram is the 26~GB/s (near-peak) PCIe x16 channel that connects \BASALISC{} to its host and carries data and instructions. We expect applications with a large working set to be performance-limited by our DDR4 bandwidth. The twin DDR4 interfaces allow us to maximize the practical throughput of the DRAM subsystem, by avoiding collisions between the PCIe-to-DRAM and FHE-to-DRAM access streams.

    \item \textbf{(Orange) The \BASALISC{} ASIC} includes JTAG I/O for testing and debuging, a RISC-V CPU to configure \BASALISC{} at startup, and the controllers and physical interfaces (PHYs) for DDR4 and PCIe. These PHYs connect to the 512-bit wide Advanced eXtensible Interface 4 (AXI4) interconnect that transfers data between the DDR, PCIe, and the CTB. Both the AXI4 and CTB operate at a target cycle time of 1~GHz. As a result, the AXI has a peak bandwidth of 32~GB/s for each endpoint connection, all running in parallel.

    \item \textbf{(Green) The \BASALISC{} FHE Core Processor} includes the CTB, AXI4 infrastructure, and a Traffic Control Unit (TCU). The TCU maintains a batch-queue instruction buffer to manage instruction execution in the system. 
    FHE programs are deterministic: the compiler knows in advance about the flow of instructions and data in memory. This allows the queue to be entirely compiler-managed and fairly short (128 to 1024 instructions, depending on our performance analysis).
\end{itemize}

\section{\BASALISC{} Processing Elements}
\label{sec:proc-element}

\BASALISC{}'s on-chip PEs and their connection to the CTB are shown on the far right in \autoref{fig:system}. The \BASALISC{} PEs that rely on the CTB for data are the Multiply-Accumulate (MAC) PE (used in ciphertext addition, multiplication, and for kernels of operations such as key switching); the Permutation PE (used to permute data into preferred orders to achieve NTT processing, and also used for automorphisms); and the NTT PE (used to accomplish number-theoretic transforms efficiently). We describe their capabilities below.

Whereas \autoref{fig:system} shows single PE instances, their implementation is a massively multicore architecture that exploits innate parallelism in FHE ciphertext computations. FHE arithmetic in RNS representation offers four types of parallelism: \emph{(i)} over multiple ciphertexts, \emph{(ii)} over the polynomials within a ciphertext, \emph{(iii)} over the residue levels of a polynomial, and \emph{(iv)} over the coefficients of a residue polynomial. Prior work has focused on \emph{(iii)}, instantiating multiple so-called Residue Polynomial Arithmetic Units (RPAUs) \cite{DBLP:journals/tc/TuranRV20, DBLP:journals/iacr/RoyMAKSY21,cryptoeprint:2022:480}. In contrast, \BASALISC{} focuses on exploiting \emph{(iv)}, due to two key observations. First, the number of residues decreases with the modulus level in the BGV scheme, leading to would-be idle RPAUs as the computation gets closer to bootstrapping. Second, as the lowest level of parallelism, coefficient-level parallelism offers the best opportunity to exploit locality of reference.

\subsection{Number-Theoretic Transform PE}

Because of the focus on coefficient-level parallelism, \BASALISC{} implements a high-radix NTT PE. We expect that many \BASALISC{} FHE applications will employ ring dimension $N = 65536 = 256^2$ to enable bootstrapping and thus arbitrary-depth computation. Thus, our NTT PE employs a radix-256 butterfly, allowing us to compute 65536-point NTTs with only two round trips to memory for each coefficient. NTTs of smaller sizes can be computed through shortcut paths in our NTT butterfly network.

Following the generalized Cooley-Tukey NTT description of \autoref{eq:cooley-tukey},
a radix-256 NTT chooses $N_1=N_2=256$. The main arithmetic NTT unit consists of a 256-point NTT (that computes the inner $N_1$-point NTT and outer $N_2$-point NTT) followed by 255 post-multipliers (that multiply with the twiddles $\omega_N^{n_2k_1}$). We employ a standard DIF flow graph for the 256-point NTT, where we replace multiplications $\omega^0 = 1$ with simple pipeline balancing registers. Through this optimization, the $N=256$-point sub-NTTs are implemented with only 769 modular multipliers, instead of $N/2\log(N)=1024$.

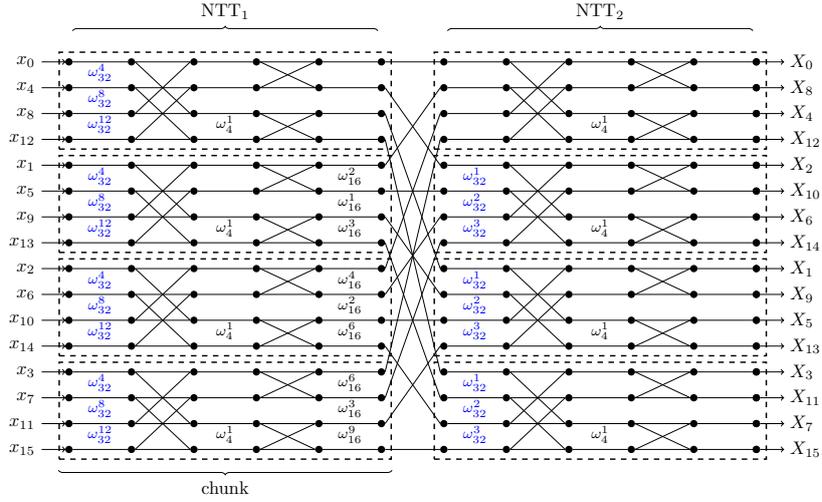
\begin{figure}
        \centering
        \resizebox{0.8\linewidth}{!}{
\newcounter{x}\newcounter{y}

\begin{tikzpicture}[yscale=0.5, xscale=1.2, node distance=0.3cm, auto]

    \foreach \y / \idx in {0/0,1/4,2/8,3/12,4/1,5/5,6/9,7/13,
                           8/2,9/6,10/10,11/14,12/3,13/7,14/11,15}
    {
        \node[n, pin={[pin edge={<-, black}]left:$x_{\idx}$}] (N--1-\y) at (-1,-\y) {};
    }
    
    \foreach \y in {0,...,15}
        \foreach \x in {0,1,2,3,4}
        {
            \node[n] (N-\x-\y) at (\x,-\y) {};
        }

    \foreach \y in {0,...,3}
        \path (N-3-\y) edge[-] node [above, midway] {\footnotesize} (N-4-\y);
        
    \foreach \y in {4,8,12}
        \path (N-3-\y) edge[-] node [above, midway] {\footnotesize} (N-4-\y);

    \path (N-3-5) edge[-] node [above, midway] {\footnotesize $\omega_{16}^2$} (N-4-5);        
    \path (N-3-6) edge[-] node [above, midway] {\footnotesize $\omega_{16}^1$} (N-4-6);        
    \path (N-3-7) edge[-] node [above, midway] {\footnotesize $\omega_{16}^3$} (N-4-7); 
        
    \path (N-3-9) edge[-] node [above, midway] {\footnotesize $\omega_{16}^4$} (N-4-9);        
    \path (N-3-10) edge[-] node [above, midway] {\footnotesize $\omega_{16}^2$} (N-4-10);        
    \path (N-3-11) edge[-] node [above, midway] {\footnotesize $\omega_{16}^{6}$} (N-4-11);        
       
    \path (N-3-13) edge[-] node [above, midway] {\footnotesize $\omega_{16}^6$} (N-4-13);        
    \path (N-3-14) edge[-] node [above, midway] {\footnotesize $\omega_{16}^{3}$} (N-4-14);        
    \path (N-3-15) edge[-] node [above, midway] {\footnotesize $\omega_{16}^{9}$} (N-4-15);  
        
    \foreach \y in {0,4,8,12}
        \path (N--1-\y) edge[-] node [above, midway] {\footnotesize} (N-0-\y);
    \foreach \y in {1,5,9,13}
        \path (N--1-\y) edge[-] node [above, midway] {\footnotesize $\color{blue}{\omega_{32}^4}$} (N-0-\y);
    \foreach \y in {2,6,10,14}
        \path (N--1-\y) edge[-] node [above, midway] {\footnotesize $\color{blue}{\omega_{32}^8}$} (N-0-\y);
    \foreach \y in {3,7,11,15}
        \path (N--1-\y) edge[-] node [above, midway] {\footnotesize $\color{blue}{\omega_{32}^{12}}$} (N-0-\y);        

    \foreach \y in {0,...,15}
        \path (N-0-\y) edge[-] (N-1-\y);
    
    \foreach \y in {0,1,2,4,5,6,8,9,10,12,13,14}
        \path (N-1-\y) edge[-] (N-2-\y);  
     
    \foreach \y in {3,7,11,15}
        \path (N-1-\y) edge[-] node [above, midway] {\footnotesize $\omega_{4}^1$} (N-2-\y); 
    \foreach \y in {0,...,15}
        \path (N-2-\y) edge[-] (N-3-\y);

    \foreach \sourcey / \desty in {0/1,2/3,4/5,6/7,
                               8/9,10/11,12/13,14/15,
                               1/0,3/2,5/4,7/6,
                               9/8,11/10,13/12,15/14}
    \path (N-2-\sourcey.east) edge[-] (N-3-\desty.west);
   
    \foreach \sourcey / \desty in {0/2,1/3,2/0,3/1,
                                       4/6,5/7,6/4,7/5,
                                       8/10,9/11,10/8,11/9,
                                       12/14,13/15,14/12,15/13}
    \path (N-0-\sourcey.east) edge[-] (N-1-\desty.west);

    \foreach \y in {0,...,15}
        \foreach \x in {5,6,7,8,9, 10}
        {
            \node[n] (N-\x-\y) at (\x,-\y) {};
        }

    \foreach \y in {0,...,15}
        \path (N-6-\y) edge[-] (N-7-\y);
    
    \foreach \y in {0,1,2,4,5,6,8,9,10,12,13,14}
        \path (N-7-\y) edge[-] (N-8-\y);  
     
    \foreach \y in {3,7,11,15}
        \path (N-7-\y) edge[-] node [above, midway] {\footnotesize $\omega_{4}^1$} (N-8-\y); 
    \foreach \y in {0,...,15}
        \path (N-8-\y) edge[-] (N-9-\y);
 
     \foreach \y in {0,...,15}
        \path (N-9-\y) edge[-] (N-10-\y);
        
    \foreach \sourcey / \desty in {0/1,2/3,4/5,6/7,
                              8/9,10/11,12/13,14/15,
                              1/0,3/2,5/4,7/6,
                              9/8,11/10,13/12,15/14}
    \path (N-8-\sourcey.east) edge[-] (N-9-\desty.west);

    \foreach \sourcey / \desty in {0/2,1/3,2/0,3/1,
                                      4/6,5/7,6/4,7/5,
                                      8/10,9/11,10/8,11/9,
                                      12/14,13/15,14/12,15/13}
    \path (N-6-\sourcey.east) edge[-] (N-7-\desty.west);

        \foreach \sourcey / \desty in {0/0,1/4,2/8,3/12,4/1,5/5,6/9,7/13,8/2,9/6,10/10,11/14,12/3,13/7,14/11,15/15}
    \path (N-4-\sourcey.east) edge[-] (N-5-\desty.west);
    
    \foreach \y in {0,...,3}
        \path (N-5-\y) edge[-] node [above, midway] {\footnotesize} (N-6-\y);
        
    \foreach \y in {4,8,12}
        \path (N-5-\y) edge[-] node [above, midway] {\footnotesize} (N-6-\y);

    \foreach \y in {5,9,13}
        \path (N-5-\y) edge[-] node [above, midway] {\footnotesize$\color{blue}{\omega_{32}^1}$} (N-6-\y);
      
    \foreach \y in {6,10,14}
        \path (N-5-\y) edge[-] node [above, midway] {\footnotesize$\color{blue}{\omega_{32}^2}$} (N-6-\y);
        
    \foreach \y in {7,11,15}
        \path (N-5-\y) edge[-] node [above, midway] {\footnotesize$\color{blue}{\omega_{32}^3}$} (N-6-\y);
        
    \foreach \y / \idx in {0/0,1/8,2/4,3/12,4/2,5/10,6,7/14,
                      8/1,9,10/5,11/13,12/3,13/11,14/7,15}    
    {
        \node[n, pin={[pin edge={->, black}]right:$X_{\idx}$}] (N-10-\y) at (10,-\y) {};
    }
    
    \draw [decorate,decoration={brace, raise=0.6cm}] (N--1-0.east) -- (N-4-0.west) node[above=0.7cm, midway] {NTT$_1$};

    \draw [decorate,decoration={brace, raise=0.6cm}] (N-5-0.east) -- (N-10-0.west) node[above=0.7cm, midway] {NTT$_2$};
    
    \node[draw=black, dashed, thick, fit=(N--1-15)(N-4-15)(N--1-12)(N-4-12)](PE) {};
    
    \node[draw=black, dashed, thick, fit=(N--1-0)(N-4-0)(N--1-3)(N-4-3)]() {};

    \node[draw=black, dashed, thick, fit=(N--1-4)(N-4-4)(N--1-7)(N-4-7)]() {};    

    \node[draw=black, dashed, thick, fit=(N--1-8)(N-4-8)(N--1-11)(N-4-11)]() {};

    \node[draw=black, dashed, thick, fit=(N-5-12)(N-10-12)(N-5-15)(N-10-15)]() {}; 
    
    \node[draw=black, dashed, thick, fit=(N-5-0)(N-10-0)(N-5-3)(N-10-3)]() {}; 
        
    \node[draw=black, dashed, thick, fit=(N-5-4)(N-10-4)(N-5-7)(N-10-7)]() {}; 
            
    \node[draw=black, dashed, thick, fit=(N-5-8)(N-10-8)(N-5-11)(N-10-11)]() {}; 

    \draw [decorate,decoration={brace, raise=0.2cm}] (PE.south east) -- (PE.south west) node[below=0.3cm, midway] {chunk};

\end{tikzpicture}}
    
       \caption{16-point radix-4 negacyclic NTT flow graph. Extra negacyclic twiddles (in blue) are decomposed into two pre-multiply passes.  \label{fig:ntt-flow-graph}}
    
\end{figure}

As discussed in \autoref{appendix:ntt}, 
additional pre- and post-multiplication steps are required to construct negacyclic forward and inverse NTTs from regular NTTs. Because a radix-256 butterfly already includes an array of 255 post-multipliers, it suffices to add 255 pre-multipliers to support negacyclic NTTs. 
\autoref{fig:ntt-flow-graph} illustrates how a radix-4 unit is composed to compute the full NTT flow graph in two passes that each take 4 chunks. In between the passes is an implicit memory transposition that we enable with a \emph{conflict-free CTB design}.

Our NTT PE instantiates four parallel 3-stage NTT units. Each unit is deeply pipelined with 40 pipeline stages in order to run at 2~GHz. Together, these four parallel pipes consume 1024 32-bit residue polynomial coefficients at that 2 GHz rate -- sufficient to consume all available data bandwidth from the CTB.

\subsubsection{Conflict-Free Schedule}

A well-known performance inhibitor for NTTs is that successive NTT passes access coefficients at different memory strides, introducing access conflicts in memory. Prior NTT accelerators present custom access patterns and reordering techniques that only work for small-radix NTT architectures \cite{DBLP:conf/ches/RoyVMCV14, riazi_heax_2020} or require expensive in-memory transpositions \cite{10.1145/3466752.3480070}. \BASALISC{} avoids reinventing the wheel, instead building upon years of DSP literature \cite{1162854,142032,DBLP:journals/tsp/Ma99}. The most high-performance FFT accelerators present \emph{conflict-free schedules} \cite{DBLP:journals/tvlsi/TsaiL11,DBLP:journals/tcas/ReisisV08,7070875} to tackle this exact issue.

Conceptually, a $N=N_1N_2=256^2$-point radix-256 NTT can be represented as a two-dimensional NTT, where the data is laid-out with $N_1=256$ rows and $N_2=256$ columns. In this format, the inner $N_1$-point NTT requires coefficients in column-major order, whereas the outer $N_2$-point NTT requires data in row-major order. The crux of building conflict-free NTT schedules is to structure the data so that it can be read out in either order without bank conflicts. This requires a minimum of 256 independently addressable banks, each containing $2^{16}$ bank addresses (for a total CTB size of $2^{24}$ values).

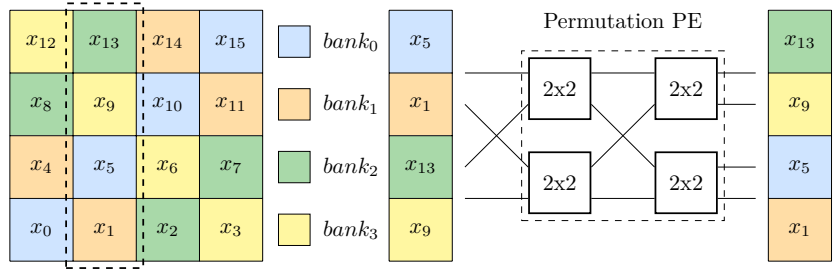
\begin{figure}
    \centering
    
\resizebox{0.8\linewidth}{!}{
  \begin{tikzpicture}[every node/.style={minimum size=1cm-\pgflinewidth, outer sep=0pt}]

\definecolor{bank0}{HTML}{CFE4FF}
\definecolor{bank1}{HTML}{FFDDA6}
\definecolor{bank2}{HTML}{A9D9A9}
\definecolor{bank3}{HTML}{FFF7A1}

    \node (x0) at (0.5,0.5) {$x_0$};
    \node (x1) at (1.5,0.5) {$x_1$};
    \node (x2) at (2.5,0.5) {$x_2$};
    \node (x3) at (3.5,0.5) {$x_3$};
    
    \node (x4) at (0.5,1.5) {$x_4$};
    \node (x5) at (1.5,1.5) {$x_5$};
    \node at (2.5,1.5) {$x_6$};
    \node at (3.5,1.5) {$x_7$};
    
    \node (x8) at (0.5,2.5) {$x_8$};
    \node (x9) at (1.5,2.5) {$x_9$};
    \node at (2.5,2.5) {$x_{10}$};
    \node at (3.5,2.5) {$x_{11}$};

    \node (x12) at (0.5,3.5) {$x_{12}$};
    \node (x13) at (1.5,3.5) {$x_{13}$};
    \node at (2.5,3.5) {$x_{14}$};
    \node at (3.5,3.5) {$x_{15}$};

    \node[fill=bank0] at (0.5,0.5) {$x_0$};
    \node[fill=bank1] at (1.5,0.5) {$x_1$};
    \node[fill=bank2] at (2.5,0.5) {$x_2$};
    \node[fill=bank3] at (3.5,0.5) {$x_3$};
    
    \node[fill=bank1] at (0.5,1.5) {$x_4$};
    \node[fill=bank0] at (1.5,1.5) {$x_5$};
    \node[fill=bank3] at (2.5,1.5) {$x_6$};
    \node[fill=bank2] at (3.5,1.5) {$x_7$};
    
    \node[fill=bank2] at (0.5,2.5) {$x_8$};
    \node[fill=bank3] at (1.5,2.5) {$x_9$};
    \node[fill=bank0] at (2.5,2.5) {$x_{10}$};
    \node[fill=bank1] at (3.5,2.5) {$x_{11}$};

    \node[fill=bank3] at (0.5,3.5) {$x_{12}$};
    \node[fill=bank2] at (1.5,3.5) {$x_{13}$};
    \node[fill=bank1] at (2.5,3.5) {$x_{14}$};
    \node[fill=bank0] at (3.5,3.5) {$x_{15}$};
    
    \filldraw [fill=bank0, draw=black] (4.25,3.25) rectangle (4.75,3.75);

    \filldraw [fill=bank1, draw=black] (4.25,2.25) rectangle (4.75,2.75);
    
    \filldraw [fill=bank2, draw=black] (4.25,1.25) rectangle (4.75,1.75);
    
    \filldraw [fill=bank3, draw=black] (4.25,0.25) rectangle (4.75,0.75);
    
    \node at (5.4,3.5) {$bank_0$};
    \node at (5.4,2.5) {$bank_1$};
    \node at (5.4,1.5) {$bank_2$};
    \node at (5.4,0.5) {$bank_3$};
    
    \node[fill=bank0] at (6.5,3.5) {$x_5$};
    \node[fill=bank1] at (6.5,2.5) {$x_1$};
    \node[fill=bank2] at (6.5,1.5) {$x_{13}$};
    \node[fill=bank3] at (6.5,0.5) {$x_9$};
    \draw[step=1cm,color=black] (6,0) grid (7,4);
    
    \node[fill=bank0] at (12.5,1.5) {$x_5$};
    \node[fill=bank1] at (12.5,0.5) {$x_1$};
    \node[fill=bank2] at (12.5,3.5) {$x_{13}$};
    \node[fill=bank3] at (12.5,2.5) {$x_9$};
    \draw[step=1cm,color=black] (12,0) grid (13,4);

    \node (rect0) at (8.7,2.75) [draw,thick] {2x2};
    \node (rect1) at (8.7,1.25) [draw,thick] {2x2};
    
    \node (rect2) at (10.7,2.75) [draw,thick] {2x2};
    \node (rect3) at (10.7,1.25) [draw,thick] {2x2};

    \draw (7.2, 3) -- (8.2, 3);
    \draw (7.2, 2.5) -- (8.2, 1.5);
    \draw (7.2, 1.5) -- (8.2, 2.5);
    \draw (7.2, 1) -- (8.2, 1);

    \draw (11.8, 3) -- (11.2, 3);
    \draw (11.8, 2.5) -- (11.2, 2.5);
    \draw (11.8, 1.5) -- (11.2, 1.5);
    \draw (11.8, 1) -- (11.2, 1);

    \draw (9.2, 3) -- (10.2, 3);
    \draw (9.2, 2.5) -- (10.2, 1.5);
    \draw (9.2, 1.5) -- (10.2, 2.5);
    \draw (9.2, 1) -- (10.2, 1);

     \node[draw=black, dashed, fit=(rect0)(rect1)(rect2)(rect2), label=above:{Permutation PE}]() {};

    \draw[step=1cm,color=black] (0,0) grid (4,4);
    \node[draw=black, dashed, thick, fit=(x1)(x5)(x9)(x13)]() {};

\end{tikzpicture}}
    \caption{Example conflict-free CTB layout for a 16-point radix-4 NTT. Data is striped using the equation $bank = row \xor col$, which ensures that both entire columns or entire rows can be read out without bank conflicts. The on-the-fly Permutation PE maps values from \emph{bank order} into \emph{natural order}, as illustrated for access to the second column.
     }

    \label{fig:conflict-free}
\end{figure}

We employ a conflict-free layout based on XOR-permutations \cite{7070875}, as illustrated in \autoref{fig:conflict-free}. In this layout, data with logical address $\{row, col\}$ is stored at $bank = row \xor col$.
This layout ensures that each unique index for every element in every row and column corresponds to a unique physically accessible bank of CTB SRAM.

When reading rows or columns from the CTB, values come out of memory in \emph{bank order}, one value for each bank from bank 0 to 255. However, operations like NTT require values in \emph{natural order}: when accessing a row, we need values sorted by column from 0 to 255, and when accessing a column, we need values sorted by row from 0 to 255. Thus, when accessing row $r$, we must map bank $i$ to index $i \xor r$. Likewise, when accessing column~$c$, we must map bank $i$ to index $i \xor c$. 

We build a custom ``on-the-fly'' Permutation PE to compute these XOR-based permutations as data moves to or from the other PEs. Furthermore, we observe a remarkable optimization opportunity for this unit. By implementing a slightly more general permutation PE that supports permutations of the form $i \mapsto (i \cdot a + b) \xor c$, we can not only use the Permutation PE to implement conflict-free XOR permutations, but also any BGV ring automorphism \emph{without additional hardware}. See \autoref{sec:perm-pe} for more details.

\subsubsection{Twiddle Factor Factory}
\label{sec:twiddlefac}

Similarly to polynomial residue coefficients, twiddle factors are 32-bit integers. There are $N$ twiddle factors for each residue for both forward and inverse NTT, and a maximum of $56$ residues at max-capacity key switching, together requiring $\sim$29.4~MB of twiddle factor material in a naive implementation. Moreover, our four NTT units have 5116 multipliers total that must be fed each cycle with twiddles, requiring massively parallel access into this storage memory. \BASALISC{} prevents this storage requirement in two ways. First, we contribute new insights and a twiddle decomposition method that reduces the required parallel number of distinct twiddle accesses. Second, we develop a custom \emph{twiddle factor factory} that drastically reduces the number of twiddles stored. In the remainder, 
we analyze only the forward NTT, but %
identical optimizations apply to the %
inverse NTT.

For a forward negacyclic NTT, each input $x_i$ is pre-multiplied by the twiddle $\phi^i = \omega_{2N}^i$. Using techniques from the DSP literature \cite{7553547}, we %
decompose the additional negacyclic twiddles to extract a regular pattern, and to distribute them evenly between the two NTT passes in the flow graph. This is illustrated in \autoref{fig:ntt-flow-graph} by the extra twiddles %
in blue. The benefit of this technique is twofold. Firstly, it can be seen that the pre-multiplications become identical for each chunk in both passes. This allows the four NTT units to share the same pre-multiply twiddles, and drastically reduces the total number of pre-multiply twiddles from $N=256^{2}$ to $2 \cdot \sqrt{N}$, easily fitting in a small SRAM. Second, the internal butterfly twiddles (powers of $\omega_{256}$) are now a strict subset of the pre-multiply twiddle in the first pass (powers of $\omega_{512}$). Both can therefore be routed from the same small SRAM. 

The remaining twiddle factor complexity sits in the post-multiply twiddles. For each chunk $k$, there are 255 twiddles $\omega_{256^2}^{ik}$. An SRAM storing vectors of 255 twiddles with depth 255 for each residue is still much too large. We propose a technique to reduce the \emph{width} of this SRAM. It can be coupled with techniques that reduce the \emph{depth} of this SRAM, such as On-the-fly-Twiddling (OT) \cite{DBLP:conf/iiswc/KimJPA20}. To reduce the width, we propose  a \emph{power generator circuit} that trades SRAM storage for multipliers. The main idea is as follows. By using the identity $\omega_{256^2}^{ik} = \omega_{256^2/k}^i$, it can be observed that the required twiddles for chunk~$k$ are always the 255 consecutive powers of a \emph{seed} value $\omega = \omega_{256^2/k}$. Using only~$\omega$, we can compute its successive powers in a number of multiply layers. The first layer computes $\omega^2$ from $\omega$, with a single multiplier. The second layer takes $\omega^2$ and $\omega$ to compute $\omega^4$ and $\omega^3$, and so forth. Every multiplier in the circuit produces a unique value that is used as an output, so the number of multipliers to generate 255 powers from~$\omega$ is simply 254. Using this technique, instead of storing vectors of 255 twiddles with depth 255 for each residue, it suffices to store the single seeds with depth 255.

\subsection{Permutation PE}

\label{sec:perm-pe}

A pair of Permutation PEs forms the interface between the CTB and the other PEs. Our original contribution is a slightly more generalized Permutation PE that can support both conflict-free schedules required by NTT operations, as well as BGV automorphisms \emph{with the same hardware}. In order to do so, the Permutation PE is generalized to compute permutations of the form $i \mapsto (i \cdot a + b) \xor c$. Each permutation unit reorders an array of input coefficients to produce a permuted output array of the same length. 

The \emph{Read Permutation PE} unscrambles data in conflict-free CTB bank ordering in order to pass it to the other PEs expecting natural ordering. It is a specialized instance of the more general Permutation PE that only implements permutations $i \mapsto i \xor c$, requiring values $a = 1$ and $b= 0$. The \emph{Write Permutation PE} passes data in the opposite direction. It implements the general permutation $i \mapsto (i \cdot a + b) \xor c$ in order to re-scramble the data into its conflict-free layout, or to compute ring automorphisms. In the latter case, the output of the Read Permutation PE is fed directly into the input of the Write Permutation PE to achieve the complete operation of the automorphism.

Each Permutation PE itself is split into two portions. Firstly, the data-permutation portion of the logic is implemented using $2\times 2$ switch nodes placed using an Omega-Network topology. Secondly, a configuration portion takes constants $a$, $b$, and $c$ in order to generate the routing pattern for the switches in the network. The configuration portion of the logic attaches the routing pattern to the data and the combined payload word is sent through the network. The switch nodes forward the data according to the least significant bit of the pattern part of the payload data, which is also removed before forwarding. 
Thus the message is reduced by one bit at each stage of the network, and at the end, the payload only contains the data portion.

\subsection{Multiply-Accumulate PE} 
\label{sec:mac-pe}

\begin{wrapfigure}{r}{0.52\linewidth}
    \centering
    \vspace{-0.9cm}
    \includegraphics[width=\linewidth]{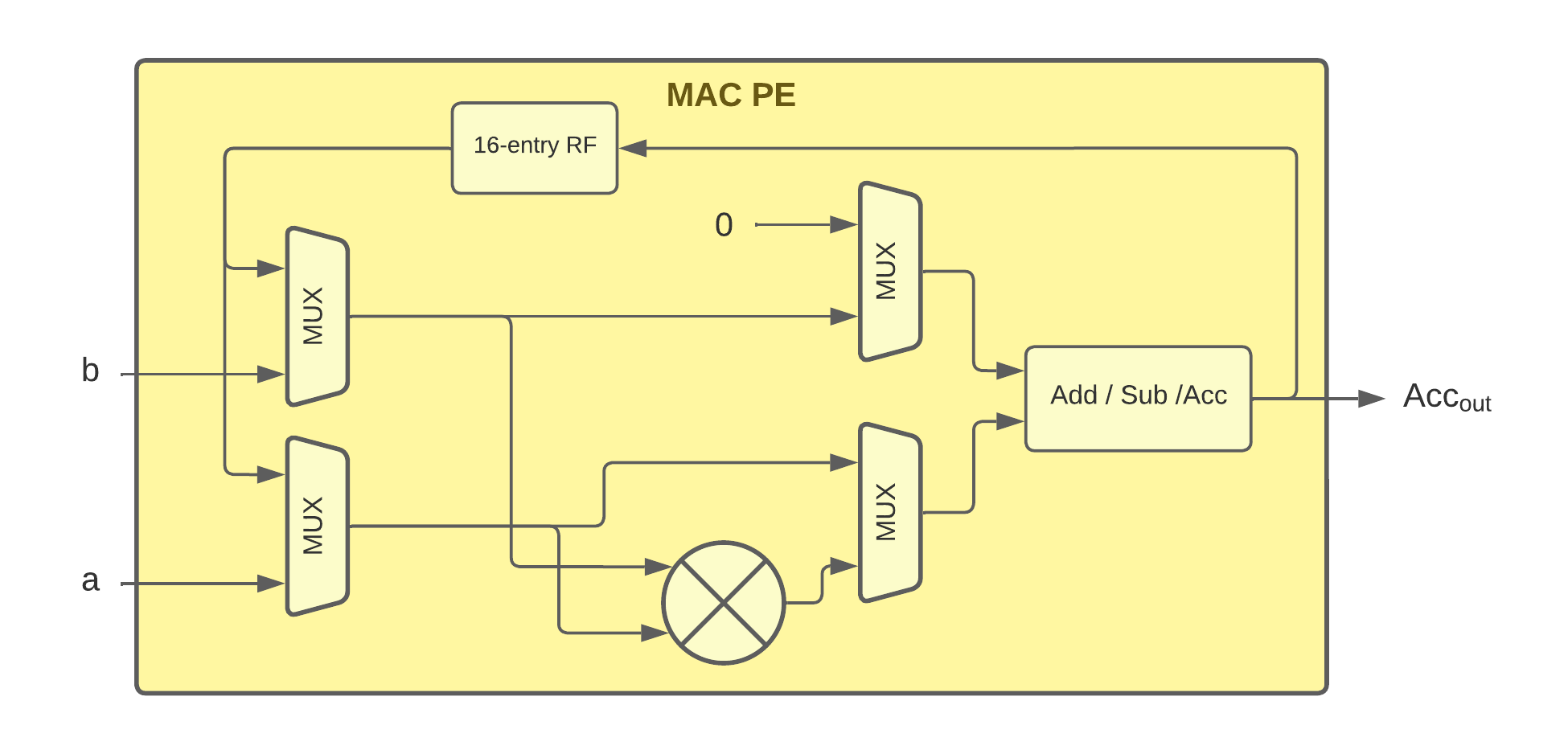}
     \caption{Simplified Diagram of MAC PE architecture.}
    \label{fig:mac}
\end{wrapfigure}
We realize modular addition and multiplication for FHE in the Multiply-Accumulate (MAC) PE, shown in \autoref{fig:mac}. This pipelined unit can start 2048 32-bit modular addition or modular multiply operations each cycle, if data is available. Because the MAC PE is built with asynchronous logic, it free-runs at 1.6~GHz when not accessing the 1~GHz CTB.
Therefore, operations that read and/or write to the CTB are limited by the 1~GHz CTB bottleneck, while other operations that operate on local data (accumulator register or register file) can accelerate to 1.6~GHz, without using additional logic. The asynchronous logic provides significant area and latency savings, compared to wide Clock Domain Crossings (CDCs) that would be necessary to achieve this 60\% frequency increase using a clocked approach instead.
At left in the figure, the 2048 $a$ inputs - each 32-bit in size - come from the CTB. The $b$ inputs are replicated copies of a 32-bit constant from the instruction stream. The MAC has a 16-entry Register File (RF), shown at top in the figure. In addition, 
there is a single
accumulator register at the output of the adder/subtractor/accumulator unit, shown at right in the figure.

Using the multiplexers shown in the figure, 
this arrangement can accomplish a variety of functions. Residue chunk multiplication by or addition of a constant to each coefficient can be accomplished at full rate: 2048 32-bit operations per cycle. Multiplication or addition of chunks when both are sourced directly from the CTB can be accomplished at half-rate, using a register to buffer one operand from the CTB, and directly feeding the second operand into the operation from the CTB in a second read cycle. Acceleration of tight kernels that repeatedly process the same chunks can be achieved by storing up to 16 different chunks in the RF and operating on them at full rate. Finally, there is a multiply-add capability that allows double-rate processing: a multiply and accumulate in every cycle. The above possibilities are impacted by the needed write bandwidth to the CTB. Write operations might occur as often as for every chunk result, or much less often when the local RF or the accumulator store results during tight kernel operations.

A particularly important example of kernel acceleration in the MAC is key switching, %
which typically makes up the large majority of the workload of an FHE program.
We implement the so-called \emph{hybrid key switching} algorithm~\cite{kpz21}, which uses the ``fast base extension'' subroutine from \autoref{eq:fast-base-extension} as an inner loop. For the parameters of \autoref{tab:params}, fast base extension first pre-computes a table of 12 residue polynomials, and then computes 40 different weighted sums of those values, with constant weights. Use of the local registers in the MAC PE and the compound multiply-accumulate function realizes a 44× improvement compared to a naive design. In addition, this approach reduces the CTB usage from nearly 100\% to only 10.6\%, saving 90\% of the CTB for use by the other PEs.

\subsection{Modular Multiplier Arithmetic Optimization}
\label{sec:mont-mul-design}

{Both the MAC and NTT butterfly units use Montgomery modular arithmetic, optimized
for NTT-friendly primes (see Mert et al.~\cite{DBLP:conf/dsd/MertOS19}). %
Specifically, instead of supporting the
\parfillskip=0pt\par}
\begin{wraptable}{r}{0.5\linewidth}
    \centering
        \caption{Area and power comparison of NTT single-butterfly unit with original and optimized Montgomery multipliers at 1~GHz.}
    \label{tab:basalisc-mult-qor}
    
\small
\vspace*{3pt}
\resizebox{.5\textwidth}{!}{
\begin{tabular}{l|c|c}
    Mult. Design & Area & TDP @0.72V, 125C \\ \hline
    Unoptimized &  3768~$\mu$m$^2$ & 7.2~$\mu$W \\ \hline
    Optimized &  2052~$\mu$m$^2$ & 4.3~$\mu$W
\end{tabular}}
\end{wraptable}
\noindent full 32-bit prime value, the multiplier is optimized to only support a subset %
where the lower 17 bits of the prime are fixed (bits 16:1 are tied to 0 and bit 0 is tied to~1). This optimization %
saves 46\% in area and 40\% in power consumption compared to a generic multiplier that can support all moduli, and is enabled by our NTT-friendly bootstrapping approach from \autoref{sec:harwareBootstrap}. The results are summarized in \autoref{tab:basalisc-mult-qor}. %

\section{\BASALISC{} Compilation and Simulation Tools}
\label{sec:compileandsim}

A major component of \BASALISC{} is co-design of software
and hardware with the intent of realizing optimal performance.
Our software
tools include a domain-specific compiler Artemidorus and a
simulator Simba, working on multiple levels of abstraction.

\subsection{Artemidorus}
Our toolchain begins with a high-level Domain-Specific Language (DSL) that allows programmers to create FHE applications to execute on \BASALISC{}, and which features data types including fixed-point numbers, vectors, and matrices. The program passes through several stages in our compiler Artemidorus.

Programs written in the Artemidorus DSL are first translated into high-level FHE circuits. The toolchain is able to type-check these circuits and combinations of circuits, to make sure the overall computation is well-formed. From this circuit representation, Artemidorus infers statistics such as circuit depth, number of bit operations, and so on.
Next, these circuit statistics are used to expand vector and matrix operations into BGV primitives; key and modulus switching operations are inferred by the tool; each operation is tagged with the length of its modulus chain (i.e., the number of prime factors in $q$), and then expanded into primitive operations on individual residue polynomials for each factor in the modulus. Finally, using a cycle-accurate model of the \BASALISC{} microarchitecture, Artemidorus allocates memory regions and registers and schedules instructions. Instruction traces are produced at our three levels of the ISA, which pass to Simba for performance or correctness simulation. Especially \emph{Simba-micro}, the micro-level performance simulator, %
is essential to evaluate \BASALISC{} at this point in the design stage. %

\subsection{Simba-micro}
Our micro-level performance simulator employs a step-based operational semantics to model the execution of the \BASALISC{} coprocessor. There are five basic operational components: the CTB, and the four PEs (MAC, Read Permutation, NTT, and Write Permutation). The simulator models a micro-instruction's life cycle from instruction dispatch, to data transfer from CTB to the appropriate functional unit, to proceeding down the pipeline, to the ``writeback'' phase.

In order to account for the different clock rates of the different components (see \autoref{tab:basalisc-pe-perf}), we use a global ``micro-clock'' which operates at 6 GHz  as the time increment for the model's step function. We made the simplifying assumption that the MAC operates at 1.5~GHz. In this way, we were able to model each PE's progress by causing the CTB to be accessible every 6 micro-cycles, the MAC every 4 micro-cycles, and the permutation/NTT units every 3 micro-cycles. This behavior is modeled by supplying each component with a wait counter that is reset to these values every time it is accessed; the component is only accessible if the counter is 0. If a component is accessible but has no work to do in the given micro-cycle,
it simply waits.

Each individual PE is modeled as a pipeline with a certain number of stages and ``stage capacity'' (number of coefficients that fit in each pipeline stage). The MAC's stage capacity is 2048 coefficients, while the other three have a capacity of 1024. Every time the given PE is enabled (its wait counter is 0), the pipeline advances. When there is a write at the end of the pipeline, it stays at the end of the pipeline until the CTB is available for writing. Once a pipeline is full, instruction dispatch is no longer possible to that pipeline, and the control mechanism allows the pending writes to occur.  

After execution, the following data is reported by Simba-micro: number of CTB cycles (i.e., 6 micro-cycles) of execution, overall CTB utilization (percentage of time spent reading/writing/stalling), and utilization of each PE (how ``full'' the pipelines are). %

\section{Evaluation of \BASALISC{}}
\BASALISC{} is a closed-source architecture with an implementation-in-progress but not delivered to silicon yet. 
For that reason, we evaluate \BASALISC{}
in diverse ways at this point in the design cycle. 

\subsection{Physical Realizability}

\begin{table}
\centering
    \caption{Performance characteristics of \BASALISC{} hardware elements.}
\label{tab:basalisc-pe-perf}
    
\small
\vspace*{3pt}
\begin{tabular}{c|r|r|r|r}
        Component  & $f_{max}$ & Area & TDP @0.72V & Throughput \\ \hline
     CTB$^\dagger$ &  1.0~GHz & 77.9~mm$^2$ & 9~W & 2 $\times$ 32~Tb/s\\
    MAC PE & 1.6~GHz & 7.17~mm$^2$ & 18.6~W & 102~Tb/s\\
     NTT PE & 2.0~GHz & 16~mm$^2$ & 24.6~W & 32~Tb/s\\
    Permuation PE & 2.0~GHz & 0.16~mm$^2$ & $\sim$0~W & 32~Tb/s \\ \hline
    
    PCIe & 500~MHz & 12~mm$^2$ & \multirow{2}{*}{5~W} & 26~GB/s\\
    DDR & 800~MHz & 18.2~mm$^2$ & & 51~GB/s \\
    
     \hline \hline
    \textbf{Overall} & $> 1.0$~GHz & 150~mm$^2$ & 57.5 - 115~W & N/A
\end{tabular}

\footnotesize{$^\dagger$Operation of PEs above the CTB's frequency is advantageous when each can run independently.} %
\end{table}

One way to evaluate the design of \BASALISC{} and the \BASALISC{} architecture is by a physical design implementation in a practical semiconductor process, with a reasonable target die size and operational frequency target. One resulting evaluation criterion that can be objectively measured using this approach is timing closure -- the verification that, with placement and routing of key blocks complete, and using industry best practice estimation of inter-block wire delays based on a mature floorplan, the design achieves a target operating frequency that yields useful levels of performance. In the case of \BASALISC{}, we completed placement and routing of the novel circuitry - our PEs - and the CTB RAM block. Our operational frequency target was a minimum of 1.0~GHz at the standard ``slow-slow'' (SS) process corner and a supply voltage of 0.72V in the 12nm low-power Global Foundries process. We achieved timing closure for the diverse functional units at the frequencies given in \autoref{tab:basalisc-pe-perf}. 

\begin{figure}
    \centering
    \includegraphics[width=0.8\linewidth]{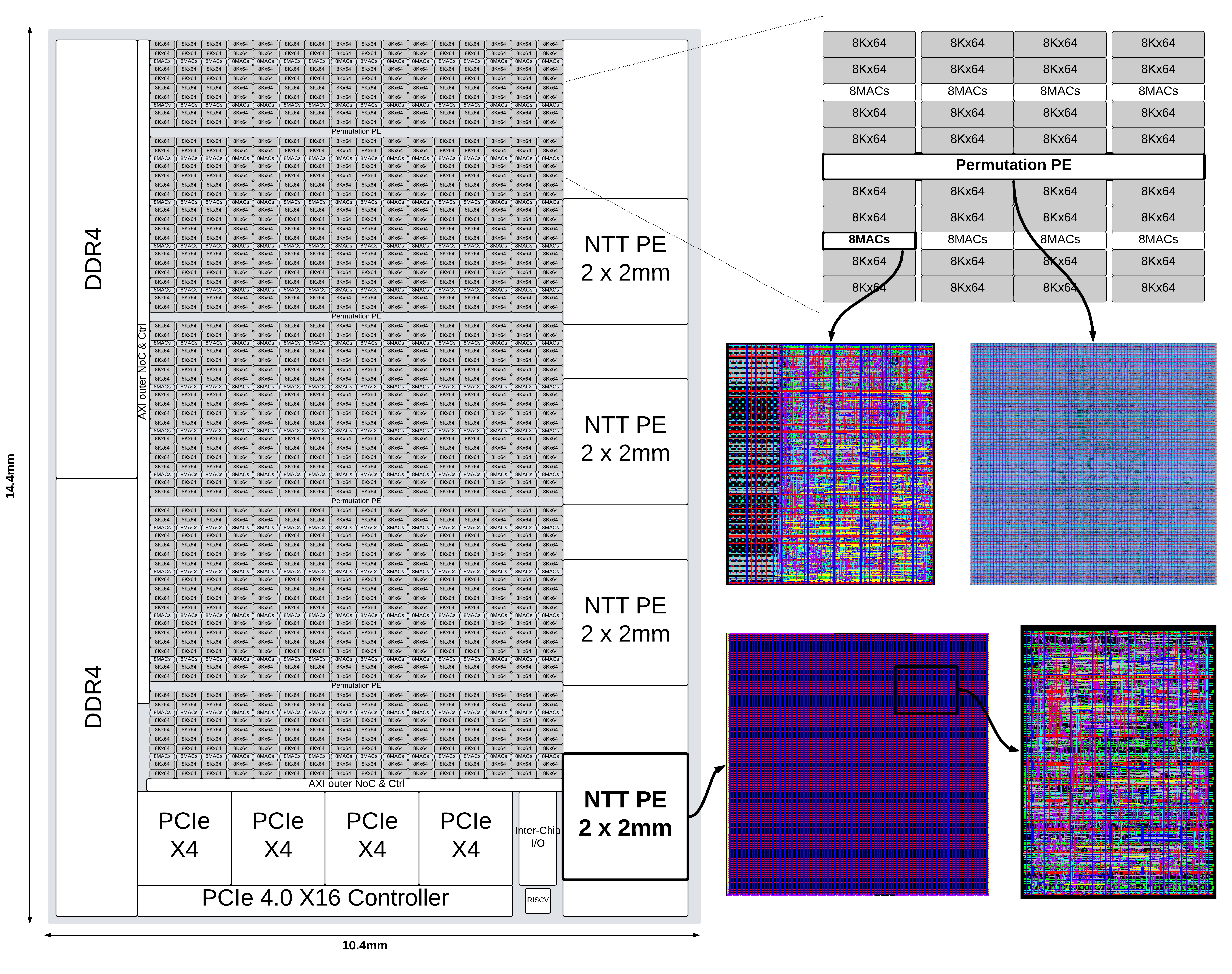}
    \caption{Floorplan of \BASALISC{} with all cells placed and intra-block routing complete. Note that the MAC PE and Permutation PE are interleaved within the CTB.}
    \label{fig:floorplan}
\end{figure}

Our floorplan shown in \autoref{fig:floorplan} uses actual IP block sizes for DDR4 DRAM, PCIe, our RAM array, and placed and routed PEs. I/O pads are part of the DDR4/PCIe macros, including bumps for power and ground. Interface clock trees run along provisioned routing channels; other big clock trees are avoided through asynchronous design. Everything is drawn assuming 75\% density, with a $200{\mu}$m peripheral gap accounting for process DRC rules, including crackstop, corners, and ESD protection. Our target die size is constrained to 150mm$^2$ with an aspect ratio of $2:1$ or less~\cite{rondeau2020data}, which we achieve with a 14.4mm~$\times$ 10.4mm layout that satisfies both of those constraints.

\subsection{Logic Emulation and Formal Verification}

A commonplace verification step prior to ASIC manufacturing provides yet another evaluation criterion: successful \emph{hardware emulation} of critical logic in the design. In the case of \BASALISC{}, that critical logic is the set of processing elements (MAC, permutation, and NNT PEs). We successfully emulated each PE in full, using test vectors extracted from our Verilog testbenches and our formal models of each PE. Each PE passed its emulation test vector suite.

In addition to hardware emulation, \BASALISC{} employs formal methods with two primary goals: first, that the design be proven mathematically correct, and second, that the design be proven consistent at every intermediate representation by demonstrating proof of equivalence. For both the mathematical and consistency proofs, \BASALISC{} employs the Cryptol language \cite{lewis2003cryptol} and related tools. 
In order to satisfy mathematical correctness, top-level FHE algorithms are expressed as a mathematical model in Cryptol.
Subsequently, using Cryptol's proof capabilities, the mathematical model is proven to sustain a set of separately-developed correctness definitions. Proof of equivalence is provided through a two-step approach. First, formal equivalence is proven between the high-level mathematical Cryptol model and a low-level logic-oriented Cryptol description using SAW \cite{10.1145/2527269.2527277}. Next, the low-level Cryptol is converted to Verilog that we prove equivalent to the optimized implementation-Verilog using the commercial Synopsys Formality tool.

\subsection{Benchmark Performance Simulation}

We evaluate \BASALISC{}'s performance on a set of benchmarks.
\autoref{tab:comparison-helib} summarizes the results and compares to an HElib software reference, which was executed on an Intel Xeon E5-2630 v2 CPU at 2.6~GHz with a single thread.
All results are generated using the
parameter set from \autoref{tab:params} (only the plaintext modulus for database lookup is chosen differently for comparison reasons). This results in ciphertext size 21 MB and 
key switching keys of size 56 MB.\footnote{Each key switching key is a matrix in $\ring_{QP}^{d \times 2}$, where the benchmarks use $d=4$. This results in a key size of
$2 d \cdot N \cdot \log_2(QP) \approx 112$ MB. However, the second column is uniformly distributed, so instead of storing it, we keep a seed and generate it on the fly. 
This technique is standard and also used in
HElib~\cite{halevi2020design}.}

The first part of \autoref{tab:comparison-helib} compares \BASALISC{} performance to HElib for a set of \emph{micro-benchmarks}: a ciphertext NTT and the basic and auxiliary homomorphic operations. Each operand is a freshly encrypted ciphertext.
We achieve major speedups for all homomorphic operations. In particular, we accelerate key switching - the most time-intensive operation - by a factor of $2.0 \cdot 10^3\times$.

\begin{table}[!htb]
    \parbox{1\linewidth}{
    \centering
    \caption{Performance comparison of HElib and \BASALISC{}.     \label{tab:comparison-helib}}
    \small
    \vspace*{3pt}
    \begin{tabular}{c||r|r|r}
        Operation & HElib
        & \BASALISC{} & Speedup \\ \hline\hline
        NTT & 27 ms & 11 $\mu$s  & $2.5 \cdot 10^3\times$ \\
        Add/Sub & 4 ms & 8 $\mu$s & $5.0\cdot 10^2\times$ \\

        Plaintext mul & 44 ms & 5 $\mu$s & $8.8 \cdot 10^{3}\times$ \\
        
        Mul (no key switch) & 58 ms  & 20 $\mu$s & $2.9 \cdot 10^3\times $ \\

        Permutation (no key switch) & 12 ms & 11 $\mu$s  & $1.1 \cdot 10^3\times$  \\

        Key switching & 580 ms & 292 $\mu$s & $2.0 \cdot 10^3\times$ \\ \hline
        Database lookup$^\dagger$ & 2,325 s & 267 ms & $8.7 \cdot 10^3\times$ \\
        Thin bootstrapping & 160 s & 40 ms & $4.0 \cdot 10^3\times$ \\
        Logistic regression & 217,000 s & 40,500 ms & $5.4 \cdot 10^3\times$
    \end{tabular}}
    \centering\footnotesize{$^\dagger$Simulation done with $t=241$ to make the result comparable with F1~\cite{10.1145/3466752.3480070}.}
\end{table}

We also simulated execution time for three realistic \emph{macro-benchmarks}: a database lookup, a bootstrapping operation and a single iteration of encrypted logistic regression training. %
The last two benchmarks require bootstrapping. This is done with HElib's thin bootstrapping procedure~\cite{halevi2021bootstrapping}, %
but adapted to our NTT-friendly approach. %

\begin{itemize}
    \item \textbf{Database lookup.} This application comes from the HElib repository~\cite{DBLookup}. An encrypted key is sent from a client to a server, then the server compares it homomorphically against a database of encrypted key-value pairs. Finally, the corresponding value is returned to the client in encrypted format.
    We achieve a speedup of 8,700 times over HElib for the server computation.
    \item \textbf{Thin bootstrapping.} This benchmark is evaluated with bootstrapping parameter $e = 4$ (see \autoref{sec:harwareBootstrap}).
    Bootstrapping requires in total 26 key switching keys, which are loaded from DRAM into the CTB during the procedure.
    Simba-micro reports only 40ms of simulated execution time, which is a speedup of 4,000 times over HElib. %
    The required noise budget before bootstrapping is 47 bits to account for a linear transformation and correctness of decryption. The remaining noise budget after bootstrapping is 790 bits, which gives 18 multiplicative levels between successive bootstrapping operations.
    \item \textbf{Logistic regression.} We estimate execution time on a single iteration of secure logistic regression (LR) training, using the 1-bit
    gradient descent algorithm of Chen et al.~\cite{chen2018logistic}. This application homomorphically trains a machine learning model on a 1,024-sample, 10-feature infant mortality data set from the US Centers for Disease Control. %
    As a \BASALISC{} instruction trace, logistic regression is composed of 900K high-level, 800M mid-level, and 27B micro-level instructions, which includes in total 513 bootstrapping operations. %
    Simba-micro reports a simulated execution time of 40.5s.
    Because logistic regression is not %
    in HElib, software execution time is estimated by counting individual operations.
    Since this benchmark is bootstrapping-dominated, the obtained speedup is similar to thin bootstrapping.
\end{itemize}

\BASALISC{} is optimized for the default %
ring dimension $N = 2^{16}$, which enables bootstrapping with a 128-bit security target. Smaller parameter sets are insufficient to %
support bootstrapping at this security level, but they can be useful in leveled applications of low multiplicative depth.
\BASALISC{}'s speedup tends to be somewhat smaller for smaller ring dimension, but it is still significant. For example, database lookup is $2,600\times$ faster for $N=2^{14}$, and %
thin bootstrapping with $N=2^{15}$ achieves $2,500\times$ speedup. %

\subsection{Related Work}

We perform a comparison to prior and concurrent FHE accelerators. %
Many early works do not report bootstrapping benchmarks or simply do not support it~\cite{poppelmann_accelerating_2015,cryptoeprint:2022:480,chen_high-speed_2015,doroz_accelerating_2015,sinha_roy_fpga-based_2019,sinha_roy_hepcloud:_2018,sinha_roy_modular_2015,riazi_heax_2020,DBLP:journals/tc/TuranRV20,DBLP:conf/fccm/AgrawalBK20,DBLP:journals/access/SuYYT20}. These architectures support only unrealistically small parameter sets, often allowing them to fully compute on-chip. Furthermore, not all accelerators implement full FHE computations, but rather individual operations such as the NTT. As one outcome, these other approaches require frequent interaction with a host processor to sequence operations and combine results. In this category of accelerators, HEAX \cite{riazi_heax_2020} achieves the most siginificant acceleration numbers, in the order of 200$\times$ for high-level operations such as key switching, compared to SEAL \cite{sealcrypto}.

Other accelerators that support bootstrapping implement the homomorphic scheme CKKS. Although CKKS and BGV
are very similar, there are important low-level differences, and an accelerator for one scheme may not necessarily support the other. \BASALISC{}'s comparison to the related BGV/CKKS accelerators is summarized in \autoref{tab:comparison-asic}.

\begin{table}
    \centering
    \caption{Comparison of \BASALISC{} with different BGV/CKKS accelerators.}
    \label{tab:comparison-asic}
    \vspace*{3pt}
    \begin{tabular}{c|c|c|c|c}
    \hline \hline
       & \textbf{\BASALISC{}}  & F1 & BTS & CraterLake  \\ \hline 
     Scheme & \textbf{BGV} & BGV/CKKS & CKKS & CKKS \\  \hline
  Area & \textbf{150mm$^2$} & 150mm$^2$ & 374mm$^2$ & 472mm$^2$ \\
  Technology & \textbf{12nm} & 12/14nm & 7nm & 12/14nm\\
  Power & \textbf{115W} & 180W & 163W & 320W \\ \hline
  Bootstrapping speedup & \textbf{4,000$\times$} & 1,830$\times$/1,195$\times$ & 2,237$\times$ & 4,400$\times$ \\
  LR speedup$^\dagger$ & \textbf{5,400$\times$} & \phantom{0000}---/7,200$\times$ & 1,306$\times$ & 2,978$\times$ \\ \hline \hline
    \end{tabular}
    \footnotesize{$^\dagger$Using \cite{chen2018logistic} for BGV LR and \cite{DBLP:conf/aaai/HanHC019} for CKKS LR.}
\end{table}

BTS~\cite{kim_bts_2021} is a CKKS accelerator with a large 374mm$^2$ area budget ($2.5\times$ ours). BTS uses a grid-based microarchitecture that lays out 2,048 PEs as 32 by 64. Conceptually, this architecture is much more complex than the simple vector architecture of \BASALISC{}. Each PE unit has a local SRAM memory, an NTT unit, a ``base extension'' unit, adders, and multipliers. This incurs a lot of communication between the PEs, so to simplify the data movement management, they treat the entire output of each PE as a ``package of coefficients''. This restricts the automorphisms that BTS can evaluate to %
the shape $c(\var) \mapsto c(\var^{5^i})$. %
While sufficient for CKKS, BTS cannot compute all automorphisms for BGV bootstrapping.
BTS uses On-the-fly Twiddling (OT) \cite{DBLP:conf/iiswc/KimJPA20} to store twiddle factors.
As we mentioned in \autoref{sec:twiddlefac}, 
this technique can be further combined with our more efficient twiddle factor factory to drastically reduce the requirements of twiddle factor storage even more. Even at the larger area budget, BTS reports similar speedups to \BASALISC{}. Respectively, in a first benchmark involving ciphertext multiplication and bootstrapping, and a second logistic regression training benchmark, BTS outperforms the Lattigo software
library by 2,237$\times$ and 1,306$\times$.

The architecture that is closest to ours and also supports BGV is F1~\cite{10.1145/3466752.3480070}. F1 is an ASIC architecture targeting the same die size (150mm$^2$), technology node (12nm~GF) and clock frequency (1~GHz). %
F1 reports a bootstrapping time of 2.4ms, compared to~40ms for our macro-benchmark. However, these numbers are not directly comparable: F1 provides lower security ($4\times$ smaller ring dimension $N$) and bootstraps a plaintext space of only 1 bit with no packing, whereas our benchmark %
has plaintext modulus $127^3$ with packing capability. %
The authors also report execution time for database lookup, but again with smaller ring dimension. F1's speedup for this application - around $7,000\times$ - is comparable to ours.
Although F1 is programmable and could support packed bootstrapping, %
it does not have enough multiplicative levels available to run it at a realistic security level.
Moreover, F1 scales poorly to larger parameter sets: it is optimized for simple BV key switching, which is less efficient than hybrid key switching for high-depth computations~\cite{kpz21}. Our MAC PE with local RF is essential to accelerate hybrid key switching. %

A novel aspect of our accelerator is the conflict-free CTB and NTT, with the corresponding Permutation PE that reuses the same hardware for NTT and automorphism. Compared to F1’s FFT algorithm, we avoid an expensive matrix transpose unit by computing the same transpose directly within the CTB. %
F1's matrix transpose unit must buffer entire ciphertext polynomials within the NTT PE, which is %
prohibitive for our parameter set. Whereas \BASALISC{} includes a highly-optimized twiddle factor factory, F1 does not describe how to implement its large twiddle factor SRAM. 

A recently proposed follow-up work to F1 is CraterLake \cite{DBLP:conf/isca/SamardzicFKMGDE22}. CraterLake makes two %
improvements over F1 that we considered from the onset, but which \BASALISC{} supports %
more efficiently. Firstly, CraterLake adds support for hybrid key switching through a complex ``vector chaining'' technique. %
\BASALISC{} supports the same algorithm efficiently within its simple MAC unit with an integrated 16-entry register file. Secondly, CraterLake observes that F1's SRAM NTT matrix transpose does not scale to larger parameters sets. It therefore decomposes this into a smaller intra-lane-group SRAM transpose, and an extra inter-lane-group fixed permutation network, both distinct from the central register file. \BASALISC{} leverages the existing CTB memory infrastructure to compute transposes without additional units or memory, using its conflict-free schedule based on XOR-permutations.

CraterLake is a substantially larger chip (472mm$^2$, 320W) than \BASALISC{},
and only evaluates CKKS. It accelerates CKKS packed bootstrapping by 4,400 times, which is similar to our speedup for BGV packed bootstrapping.
At the same time, \BASALISC{} appears closer to tape-out: CraterLake does not formally verify its PEs, does not present a layout or floorplan, and includes difficult-to-manufacture IP cores such as HBM2E.

\section{Conclusion}

FHE enables new privacy-preserving applications, but its adoption is limited because of high computational costs. \BASALISC{} accelerates FHE computations by more than three orders of magnitude over CPUs, and thereby takes a step toward practical feasibility.

In contrast to prior works, \BASALISC{} supports all BGV operations, including fully-packed bootstrapping, in a single ASIC architecture. Our design includes a complete memory hierarchy, and an ISA that supports different levels of abstraction. We propose several hardware improvements in the NTT architecture, and show that its permutation unit can be generalized to compute BGV automorphisms without additional area. \BASALISC{} saves area and power consumption by restricting its multipliers to NTT-friendly primes. This optimization still allows BGV bootstrapping, so it does not compromise generality.

We evaluate the design of \BASALISC{} for correctness and performance. Our functional units are emulated and formally verified to meet their specification. We also simulate performance on three FHE macro-benchmarks, showing over 5,000 times speedup compared to classical software implementations. \BASALISC{} has been selected as a candidate for future fabrication of an IC that can be applied in real-world applications.

\section*{Acknowledgments}
This material is based upon work supported by the Defense Advanced Research Projects Agency (DARPA) under Contract No. HR0011-21-C-0034. The views, opinions, and/or findings expressed are those of the authors and should not be interpreted as representing the official views or policies of the Department of Defense or the U.S. Government. This work was additionally supported in part by CyberSecurity Research Flanders with reference number VR20192203 and the Research Council KU Leuven (C16/15/058). Michiel Van Beirendonck is funded by Research Foundation – Flanders (FWO) as Strategic Basic (SB) PhD fellow (project number 1SD5621N). Robin Geelen is funded in part by Research Foundation – Flanders (FWO) under a PhD Fellowship fundamental research (project number 1162123N).

\FloatBarrier

\bibliographystyle{alpha}
\bibliography{bibliography}

\newcommand{\etalchar}[1]{$^{#1}$}
\begin{thebibliography}{CGBH{\etalchar{+}}18}

\bibitem[ABK20]{DBLP:conf/fccm/AgrawalBK20}
Rashmi~S. Agrawal, Lake Bu, and Michel~A. Kinsy.
\newblock Fast arithmetic hardware library for rlwe-based homomorphic
  encryption.
\newblock In {\em 28th {IEEE} Annual International Symposium on
  Field-Programmable Custom Computing Machines, {FCCM} 2020, Fayetteville, AR,
  USA, May 3-6, 2020}, page 206. {IEEE}, 2020.

\bibitem[ACC{\etalchar{+}}18]{HomomorphicEncryptionSecurityStandard}
Martin Albrecht, Melissa Chase, Hao Chen, Jintai Ding, Shafi Goldwasser, Sergey
  Gorbunov, Shai Halevi, Jeffrey Hoffstein, Kim Laine, Kristin Lauter, Satya
  Lokam, Daniele Micciancio, Dustin Moody, Travis Morrison, Amit Sahai, and
  Vinod Vaikuntanathan.
\newblock Homomorphic encryption security standard.
\newblock Technical report, HomomorphicEncryption.org, Toronto, Canada,
  November 2018.

\bibitem[AHU74]{DBLP:books/aw/AhoHU74}
Alfred~V. Aho, John~E. Hopcroft, and Jeffrey~D. Ullman.
\newblock {\em The Design and Analysis of Computer Algorithms}.
\newblock Addison-Wesley, 1974.

\bibitem[ASP13]{alperin2013practical}
Jacob Alperin-Sheriff and Chris Peikert.
\newblock Practical bootstrapping in quasilinear time.
\newblock In {\em Annual Cryptology Conference}, pages 1--20. Springer, 2013.

\bibitem[BEHZ16]{bajard2016full}
Jean-Claude Bajard, Julien Eynard, M~Anwar Hasan, and Vincent Zucca.
\newblock A full rns variant of fv like somewhat homomorphic encryption
  schemes.
\newblock In {\em International Conference on Selected Areas in Cryptography},
  pages 423--442. Springer, 2016.

\bibitem[BGV14]{brakerski2014leveled}
Zvika Brakerski, Craig Gentry, and Vinod Vaikuntanathan.
\newblock (leveled) fully homomorphic encryption without bootstrapping.
\newblock {\em ACM Transactions on Computation Theory (TOCT)}, 6(3):1--36,
  2014.

\bibitem[BIP{\etalchar{+}}22]{bipps22}
Charlotte Bonte, Ilia Iliashenko, Jeongeun Park, Hilder V.~L. Pereira, and
  Nigel~P. Smart.
\newblock Final: Faster fhe instantiated with ntru and lwe.
\newblock Cryptology ePrint Archive, Report 2022/074, 2022.
\newblock \url{https://ia.cr/2022/074}.

\bibitem[CFH{\etalchar{+}}13]{10.1145/2527269.2527277}
Kyle Carter, Adam Foltzer, Joe Hendrix, Brian Huffman, and Aaron Tomb.
\newblock Saw: The software analysis workbench.
\newblock In {\em Proceedings of the 2013 ACM SIGAda Annual Conference on High
  Integrity Language Technology}, HILT '13, page 15–18, New York, NY, USA,
  2013. Association for Computing Machinery.

\bibitem[CGBH{\etalchar{+}}18]{chen2018logistic}
Hao Chen, Ran Gilad-Bachrach, Kyoohyung Han, Zhicong Huang, Amir Jalali, Kim
  Laine, and Kristin Lauter.
\newblock Logistic regression over encrypted data from fully homomorphic
  encryption.
\newblock {\em BMC medical genomics}, 11(4):3--12, 2018.

\bibitem[CGGI20]{cggi20}
Ilaria Chillotti, Nicolas Gama, Mariya Georgieva, and Malika Izabach{\`e}ne.
\newblock Tfhe: fast fully homomorphic encryption over the torus.
\newblock {\em Journal of Cryptology}, 33(1):34--91, 2020.

\bibitem[CH18]{chen2018homomorphic}
Hao Chen and Kyoohyung Han.
\newblock Homomorphic lower digits removal and improved fhe bootstrapping.
\newblock In {\em Annual International Conference on the Theory and
  Applications of Cryptographic Techniques}, pages 315--337. Springer, 2018.

\bibitem[CJL{\etalchar{+}}20]{WAHC:CJLOT20}
Ilaria Chillotti, Marc Joye, Damien Ligier, Jean-Baptiste Orfila, and Samuel
  Tap.
\newblock Concrete: Concrete operates on ciphertexts rapidly by extending tfhe.
\newblock In {\em WAHC 2020--8th Workshop on Encrypted Computing \& Applied
  Homomorphic Cryptography}, volume~15, 2020.

\bibitem[CKKS17]{ckks}
Jung~Hee Cheon, Andrey Kim, Miran Kim, and Yongsoo Song.
\newblock Homomorphic encryption for arithmetic of approximate numbers.
\newblock In Tsuyoshi Takagi and Thomas Peyrin, editors, {\em Advances in
  Cryptology -- ASIACRYPT 2017}, pages 409--437, Cham, 2017. Springer
  International Publishing.

\bibitem[CMV{\etalchar{+}}15]{chen_high-speed_2015}
Donald~Donglong Chen, Nele Mentens, Frederik Vercauteren, Sujoy~Sinha Roy, Ray
  C.~C. Cheung, Derek Pao, and Ingrid Verbauwhede.
\newblock High-{Speed} {Polynomial} {Multiplication} {Architecture} for
  {Ring}-{LWE} and {SHE} {Cryptosystems}.
\newblock {\em IEEE Transactions on Circuits and Systems I: Regular Papers},
  62(1):157--166, January 2015.

\bibitem[Coh76]{1162854}
D.~Cohen.
\newblock Simplified control of fft hardware.
\newblock {\em IEEE Transactions on Acoustics, Speech, and Signal Processing},
  24(6):577--579, 1976.

\bibitem[D{\"O}S14]{doroz_accelerating_2015}
Yark{\i}n Dor{\"o}z, Erdin{\c{c}} {\"O}zt{\"u}rk, and Berk Sunar.
\newblock Accelerating fully homomorphic encryption in hardware.
\newblock {\em IEEE Transactions on Computers}, 64(6):1509--1521, 2014.

\bibitem[DSL{\etalchar{+}}18]{8259423}
Mike Davies, Narayan Srinivasa, Tsung-Han Lin, Gautham Chinya, Yongqiang Cao,
  Sri~Harsha Choday, Georgios Dimou, Prasad Joshi, Nabil Imam, Shweta Jain,
  Yuyun Liao, Chit-Kwan Lin, Andrew Lines, Ruokun Liu, Deepak Mathaikutty,
  Steven McCoy, Arnab Paul, Jonathan Tse, Guruguhanathan Venkataramanan,
  Yi-Hsin Weng, Andreas Wild, Yoonseok Yang, and Hong Wang.
\newblock Loihi: A neuromorphic manycore processor with on-chip learning.
\newblock {\em IEEE Micro}, 38(1):82--99, 2018.

\bibitem[Gar16]{7553547}
Mario Garrido.
\newblock A new representation of fft algorithms using triangular matrices.
\newblock {\em IEEE Transactions on Circuits and Systems I: Regular Papers},
  63(10):1737--1745, 2016.

\bibitem[Gen09]{DBLP:conf/stoc/Gentry09}
Craig Gentry.
\newblock Fully homomorphic encryption using ideal lattices.
\newblock In Michael Mitzenmacher, editor, {\em Proceedings of the 41st Annual
  {ACM} Symposium on Theory of Computing, {STOC} 2009, Bethesda, MD, USA, May
  31 - June 2, 2009}, pages 169--178. {ACM}, 2009.

\bibitem[GHS12a]{gentry2012fully}
Craig Gentry, Shai Halevi, and Nigel~P Smart.
\newblock Fully homomorphic encryption with polylog overhead.
\newblock In {\em Annual International Conference on the Theory and
  Applications of Cryptographic Techniques}, pages 465--482. Springer, 2012.

\bibitem[GHS12b]{gentry2012homomorphic}
Craig Gentry, Shai Halevi, and Nigel~P Smart.
\newblock Homomorphic evaluation of the aes circuit.
\newblock In {\em Annual Cryptology Conference}, pages 850--867. Springer,
  2012.

\bibitem[GV22]{geelen2022bootstrapping}
Robin Geelen and Frederik Vercauteren.
\newblock Bootstrapping for bgv and bfv revisited.
\newblock {\em Cryptology ePrint Archive}, 2022.

\bibitem[HEl]{DBLookup}
{HElib} country lookup example.
\newblock
  \url{https://github.com/homenc/HElib/tree/master/examples/BGV_country_db_lookup}.
\newblock Accessed: 2022-12-17.

\bibitem[HHCP19]{DBLP:conf/aaai/HanHC019}
Kyoohyung Han, Seungwan Hong, Jung~Hee Cheon, and Daejun Park.
\newblock Logistic regression on homomorphic encrypted data at scale.
\newblock In {\em The Thirty-Third {AAAI} Conference on Artificial
  Intelligence, {AAAI} 2019, The Thirty-First Innovative Applications of
  Artificial Intelligence Conference, {IAAI} 2019, The Ninth {AAAI} Symposium
  on Educational Advances in Artificial Intelligence, {EAAI} 2019, Honolulu,
  Hawaii, USA, January 27 - February 1, 2019}, pages 9466--9471. {AAAI} Press,
  2019.

\bibitem[HS14]{DBLP:conf/crypto/HaleviS14}
Shai Halevi and Victor Shoup.
\newblock Algorithms in helib.
\newblock In Juan~A. Garay and Rosario Gennaro, editors, {\em Advances in
  Cryptology - {CRYPTO} 2014 - 34th Annual Cryptology Conference, Santa
  Barbara, CA, USA, August 17-21, 2014, Proceedings, Part {I}}, volume 8616 of
  {\em Lecture Notes in Computer Science}, pages 554--571. Springer, 2014.

\bibitem[HS20]{halevi2020design}
Shai Halevi and Victor Shoup.
\newblock Design and implementation of helib: a homomorphic encryption library.
\newblock {\em Cryptology ePrint Archive}, 2020.

\bibitem[HS21]{halevi2021bootstrapping}
Shai Halevi and Victor Shoup.
\newblock Bootstrapping for helib.
\newblock {\em Journal of Cryptology}, 34(1):1--44, 2021.

\bibitem[Joh92]{142032}
L.G. Johnson.
\newblock Conflict free memory addressing for dedicated fft hardware.
\newblock {\em IEEE Transactions on Circuits and Systems II: Analog and Digital
  Signal Processing}, 39(5):312--316, 1992.

\bibitem[KJPA20]{DBLP:conf/iiswc/KimJPA20}
Sangpyo Kim, Wonkyung Jung, Jaiyoung Park, and Jung~Ho Ahn.
\newblock Accelerating number theoretic transformations for bootstrappable
  homomorphic encryption on gpus.
\newblock In {\em {IEEE} International Symposium on Workload Characterization,
  {IISWC} 2020, Beijing, China, October 27-30, 2020}, pages 264--275. {IEEE},
  2020.

\bibitem[KKK{\etalchar{+}}21]{kim_bts_2021}
Sangpyo Kim, Jongmin Kim, Michael~Jaemin Kim, Wonkyung Jung, Minsoo Rhu, John
  Kim, and Jung~Ho Ahn.
\newblock {BTS}: {An} {Accelerator} for {Bootstrappable} {Fully} {Homomorphic}
  {Encryption}.
\newblock {\em arXiv:2112.15479 [cs]}, December 2021.

\bibitem[KPZ21]{kpz21}
Andrey Kim, Yuriy Polyakov, and Vincent Zucca.
\newblock Revisiting homomorphic encryption schemes for finite fields.
\newblock In Mehdi Tibouchi and Huaxiong Wang, editors, {\em Advances in
  Cryptology -- ASIACRYPT 2021}, pages 608--639, Cham, 2021. Springer
  International Publishing.

\bibitem[Lat22]{lattigo}
Lattigo v3.
\newblock Online: \url{https://github.com/tuneinsight/lattigo}, April 2022.
\newblock EPFL-LDS, Tune Insight SA.

\bibitem[LFS87]{1458134}
Kun-Shan Lin, G.A. Frantz, and R.~Simar.
\newblock The tms320 family of digital signal processors.
\newblock {\em Proceedings of the IEEE}, 75(9):1143--1159, 1987.

\bibitem[LM03]{lewis2003cryptol}
Jeffrey~R Lewis and Brad Martin.
\newblock Cryptol: High assurance, retargetable crypto development and
  validation.
\newblock In {\em IEEE Military Communications Conference, 2003. MILCOM 2003.},
  volume~2, pages 820--825. IEEE, 2003.

\bibitem[Ma99]{DBLP:journals/tsp/Ma99}
Yutai Ma.
\newblock An effective memory addressing scheme for {FFT} processors.
\newblock {\em {IEEE} Trans. Signal Process.}, 47(3):907--911, 1999.

\bibitem[MAK{\etalchar{+}}22]{cryptoeprint:2022:480}
Ahmet~Can Mert, Aikata, Sunmin Kwon, Youngsam Shin, Donghoon Yoo, Yongwoo Lee,
  and Sujoy~Sinha Roy.
\newblock Medha: Microcoded hardware accelerator for computing on encrypted
  data.
\newblock Cryptology ePrint Archive, Report 2022/480, 2022.
\newblock \url{https://ia.cr/2022/480}.

\bibitem[M{\"{O}}S19]{DBLP:conf/dsd/MertOS19}
Ahmet~Can Mert, Erdin{\c{c}} {\"{O}}zt{\"{u}}rk, and Erkay Savas.
\newblock Design and implementation of a fast and scalable ntt-based polynomial
  multiplier architecture.
\newblock In {\em 22nd Euromicro Conference on Digital System Design, {DSD}
  2019, Kallithea, Greece, August 28-30, 2019}, pages 253--260. {IEEE}, 2019.

\bibitem[PNPM15]{poppelmann_accelerating_2015}
Thomas Pöppelmann, Michael Naehrig, Andrew Putnam, and Adrian Macias.
\newblock Accelerating {Homomorphic} {Evaluation} on {Reconfigurable}
  {Hardware}.
\newblock In Tim Güneysu and Helena Handschuh, editors, {\em Cryptographic
  {Hardware} and {Embedded} {Systems} -- {CHES} 2015}, Lecture {Notes} in
  {Computer} {Science}, pages 143--163, Berlin, Heidelberg, 2015. Springer.

\bibitem[PRR17]{polyakov2017palisade}
Yuriy Polyakov, Kurt Rohloff, and Gerard~W Ryan.
\newblock Palisade lattice cryptography library user manual.
\newblock 2017.

\bibitem[RAD{\etalchar{+}}78]{rivest1978data}
Ronald~L Rivest, Len Adleman, Michael~L Dertouzos, et~al.
\newblock On data banks and privacy homomorphisms.
\newblock {\em Foundations of secure computation}, 4(11):169--180, 1978.

\bibitem[RLPD20]{riazi_heax_2020}
M.~Sadegh Riazi, Kim Laine, Blake Pelton, and Wei Dai.
\newblock {HEAX:} an architecture for computing on encrypted data.
\newblock In James~R. Larus, Luis Ceze, and Karin Strauss, editors, {\em
  {ASPLOS} '20: Architectural Support for Programming Languages and Operating
  Systems, Lausanne, Switzerland, March 16-20, 2020}, pages 1295--1309. {ACM},
  2020.

\bibitem[RMA{\etalchar{+}}21]{DBLP:journals/iacr/RoyMAKSY21}
Sujoy~Sinha Roy, Ahmet~Can Mert, Aikata, Sunmin Kwon, Youngsam Shin, and
  Donghoon Yoo.
\newblock Accelerator for computing on encrypted data.
\newblock {\em {IACR} Cryptol. ePrint Arch.}, page 1555, 2021.

\bibitem[RMD{\etalchar{+}}15]{7070875}
Stephen Richardson, Dejan Marković, Andrew Danowitz, John Brunhaver, and Mark
  Horowitz.
\newblock Building conflict-free fft schedules.
\newblock {\em IEEE Transactions on Circuits and Systems I: Regular Papers},
  62(4):1146--1155, 2015.

\bibitem[Ron20]{rondeau2020data}
Tom Rondeau.
\newblock Data protection in virtual environments ({DPRIVE}), 2020.

\bibitem[RV08]{DBLP:journals/tcas/ReisisV08}
Dionysios~I. Reisis and Nikolaos Vlassopoulos.
\newblock Conflict-free parallel memory accessing techniques for {FFT}
  architectures.
\newblock {\em {IEEE} Trans. Circuits Syst. {I} Regul. Pap.},
  55-I(11):3438--3447, 2008.

\bibitem[RVM{\etalchar{+}}14]{DBLP:conf/ches/RoyVMCV14}
Sujoy~Sinha Roy, Frederik Vercauteren, Nele Mentens, Donald~Donglong Chen, and
  Ingrid Verbauwhede.
\newblock Compact ring-lwe cryptoprocessor.
\newblock In Lejla Batina and Matthew Robshaw, editors, {\em Cryptographic
  Hardware and Embedded Systems - {CHES} 2014 - 16th International Workshop,
  Busan, South Korea, September 23-26, 2014. Proceedings}, volume 8731 of {\em
  Lecture Notes in Computer Science}, pages 371--391. Springer, 2014.

\bibitem[SEA22]{sealcrypto}
{M}icrosoft {SEAL} (release 4.0).
\newblock \url{https://github.com/Microsoft/SEAL}, March 2022.
\newblock Microsoft Research, Redmond, WA.

\bibitem[SFK{\etalchar{+}}21]{10.1145/3466752.3480070}
Nikola Samardzic, Axel Feldmann, Aleksandar Krastev, Srinivas Devadas, Ronald
  Dreslinski, Christopher Peikert, and Daniel Sanchez.
\newblock F1: A fast and programmable accelerator for fully homomorphic
  encryption.
\newblock In {\em MICRO-54: 54th Annual IEEE/ACM International Symposium on
  Microarchitecture}, MICRO '21, page 238–252, New York, NY, USA, 2021.
  Association for Computing Machinery.

\bibitem[SFK{\etalchar{+}}22]{DBLP:conf/isca/SamardzicFKMGDE22}
Nikola Samardzic, Axel Feldmann, Aleksandar Krastev, Nathan Manohar, Nicholas
  Genise, Srinivas Devadas, Karim Eldefrawy, Chris Peikert, and Daniel
  S{\'{a}}nchez.
\newblock Craterlake: a hardware accelerator for efficient unbounded
  computation on encrypted data.
\newblock In Valentina Salapura, Mohamed Zahran, Fred Chong, and Lingjia Tang,
  editors, {\em {ISCA} '22: The 49th Annual International Symposium on Computer
  Architecture, New York, New York, USA, June 18 - 22, 2022}, pages 173--187.
  {ACM}, 2022.

\bibitem[SRJV{\etalchar{+}}15]{sinha_roy_modular_2015}
Sujoy Sinha~Roy, Kimmo Järvinen, Frederik Vercauteren, Vassil Dimitrov, and
  Ingrid Verbauwhede.
\newblock Modular {Hardware} {Architecture} for {Somewhat} {Homomorphic}
  {Function} {Evaluation}.
\newblock In Tim Güneysu and Helena Handschuh, editors, {\em Cryptographic
  {Hardware} and {Embedded} {Systems} -- {CHES} 2015}, Lecture {Notes} in
  {Computer} {Science}, pages 164--184, Berlin, Heidelberg, 2015. Springer.

\bibitem[SRJV{\etalchar{+}}18]{sinha_roy_hepcloud:_2018}
Sujoy Sinha~Roy, Kimmo Järvinen, Jo~Vliegen, Frederik Vercauteren, and Ingrid
  Verbauwhede.
\newblock {HEPCloud}: {An} {FPGA}-{Based} {Multicore} {Processor} for {FV}
  {Somewhat} {Homomorphic} {Function} {Evaluation}.
\newblock {\em IEEE Transactions on Computers}, 67(11):1637--1650, November
  2018.

\bibitem[SRTJ{\etalchar{+}}19]{sinha_roy_fpga-based_2019}
Sujoy Sinha~Roy, Furkan Turan, Kimmo Jarvinen, Frederik Vercauteren, and Ingrid
  Verbauwhede.
\newblock {FPGA}-{Based} {High}-{Performance} {Parallel} {Architecture} for
  {Homomorphic} {Computing} on {Encrypted} {Data}.
\newblock In {\em 2019 {IEEE} {International} {Symposium} on {High}
  {Performance} {Computer} {Architecture} ({HPCA})}, pages 387--398, February
  2019.

\bibitem[SV14]{smart2014fully}
Nigel~P Smart and Frederik Vercauteren.
\newblock Fully homomorphic simd operations.
\newblock {\em Designs, codes and cryptography}, 71(1):57--81, 2014.

\bibitem[SYYT20]{DBLP:journals/access/SuYYT20}
Yang Su, Bailong Yang, Chen Yang, and Luogeng Tian.
\newblock Fpga-based hardware accelerator for leveled ring-lwe fully
  homomorphic encryption.
\newblock {\em {IEEE} Access}, 8:168008--168025, 2020.

\bibitem[TL11]{DBLP:journals/tvlsi/TsaiL11}
Pei{-}Yun Tsai and Chung{-}Yi Lin.
\newblock A generalized conflict-free memory addressing scheme for
  continuous-flow parallel-processing {FFT} processors with rescheduling.
\newblock {\em {IEEE} Trans. Very Large Scale Integr. Syst.},
  19(12):2290--2302, 2011.

\bibitem[TRV20]{DBLP:journals/tc/TuranRV20}
Furkan Turan, Sujoy~Sinha Roy, and Ingrid Verbauwhede.
\newblock {HEAWS:} an accelerator for homomorphic encryption on the amazon
  {AWS} {FPGA}.
\newblock {\em {IEEE} Trans. Computers}, 69(8):1185--1196, 2020.

\bibitem[Zuc18]{zucca2018towards}
Vincent Zucca.
\newblock {\em Towards efficient arithmetic for Ring-LWE based homomorphic
  encryption}.
\newblock PhD thesis, Sorbonne universit{\'e}, 2018.

\end{thebibliography}

\end{document}